\documentclass[11pt]{article}
\usepackage{amssymb, latexsym, amsmath}
\usepackage{amsfonts, graphics}
\def\R{\mathbb R}

\def\A{\mathcal{A}}
\def\p{\left\langle v\right\rangle}

\def\be{\begin{equation}}
\def\ee{\end{equation}}
\def\bea{\begin{eqnarray}}
\def\eea{\end{eqnarray}}
\def\beas{\begin{eqnarray*}}
\def\eeas{\end{eqnarray*}}
\def\supp{\mathrm{supp}\,}

\def\open#1{\setbox0=\hbox{$#1$}
\baselineskip = 0pt
\vbox{\hbox{\hspace*{0.4 \wd0}\tiny $\circ$}\hbox{$#1$}}
\baselineskip = 11pt\!}
\def\fn{\open{f}}

\newcommand{\prfe}{\hspace*{\fill} $\Box$

\smallskip \noindent}

\begin{document}
\sloppy
\newtheorem{theorem}{Theorem}[section]
\newtheorem{definition}[theorem]{Definition}
\newtheorem{proposition}[theorem]{Proposition}
\newtheorem{example}[theorem]{Example}
\newtheorem{remark}[theorem]{Remark}
\newtheorem{cor}[theorem]{Corollary}
\newtheorem{lemma}[theorem]{Lemma}

\renewcommand{\theequation}{\arabic{section}.\arabic{equation}}

\title{Stability for the spherically symmetric Einstein-Vlasov 
       system---a coercivity estimate}

\author{Mahir Had\v{z}i\'{c}\\
        Massachusetts Institute of Technology\\
        Department of Mathematics\\
        Cambridge, MA 02139-4307, U.S.A.\\
        email: hadzic@math.mit.edu\\
        \ \\
        Gerhard Rein\\
        Fakult\"at f\"ur Mathematik, Physik und Informatik\\
        Universit\"at Bayreuth\\
        D-95440 Bayreuth, Germany\\
        email: gerhard.rein@uni-bayreuth.de}

        \date{January 17, 2012}
\maketitle

\begin{abstract}
The stability of static solutions of the spherically symmetric,
asymptotically flat Einstein-Vlasov system is studied using
a Hamiltonian approach based on energy-Casimir functionals. 
The main result is a coercivity estimate for
the quadratic part of the expansion of the natural energy-Casimir 
functional about an isotropic steady state. The estimate
shows in a quantified way that this quadratic part is positive definite
on a class of linearly dynamically accessible perturbations,
provided the particle distribution of the steady state is a 
strictly decreasing function of the particle energy and provided 
the steady state is not too relativistic. This should be an essential
step in a fully non-linear stability analysis for the Einstein-Vlasov
system. In the present paper it is exploited for obtaining a linearized
stability result.
\end{abstract}
%
%
%
\section{Introduction}
\setcounter{equation}{0}
The aim of the present paper is to investigate
the stability of spherically symmetric steady states of the 
Einstein-Vlasov system against spherically symmetric perturbations.
The system describes, in the context of general relativity,
the evolution of an ensemble of particles which
interact only via gravity. Galaxies or globular clusters, 
where the stars play the role of the particles, can be modeled as such
ensembles, since collisions among stars are sufficiently rare to 
be neglected. 
The particle distribution is given by a density
function $f$ on the tangent bundle $TM$ of the spacetime manifold $M$.
We assume that all particles have the
same rest mass which is normalized to unity.
Hence the particle distribution function is supported
on the mass shell
\[
PM = \{ g_{\alpha \beta} p^\alpha p^\beta = -c^2\ 
\mbox{and}\ p^\alpha\ \mbox{is future pointing}\} 
\subset TM.
\]
Here $g_{\alpha \beta}$ denotes the Lorentz metric on the 
spacetime manifold $M$
and $p^\alpha$ denote the canonical momentum coordinates
corresponding to a choice of local coordinates
$x^\alpha$ on $M$; Greek
indices always run from $0$ to $3$, and we have a specific reason for 
making the dependence on the speed of light $c$ explicit.
We assume that the coordinates are chosen such that
\[
ds^2 = c^2 g_{00} dt^2 + g_{ab} dx^a dx^b
\] 
where Latin indices run from $1$ to $3$
and $t=x^0$ should be thought of as a timelike coordinate.
On the mass shell, $p^0$ can be expressed by the remaining coordinates,
\[
p^0 = \sqrt{-g^{00}}\sqrt{1+ c^{-2} g_{ab}p^a p^b},
\]
and $f=f(t,x^a,p^b)\geq 0$. The Einstein-Vlasov system
now consists of the Einstein field equations
\begin{equation} \label{einst_gen}
G_{\alpha\beta} = 8 \pi c^{-4} T_{\alpha\beta}
\end{equation}
coupled to the Vlasov equation
\begin{equation} \label{vlasov_gen}
p^0 \partial_t f + p^a \partial_{x^a} f - 
\Gamma^a_{\beta \gamma} p^\beta p^\gamma \partial_{p^a} f = 0
\end{equation}
via the following definition of the energy momentum tensor:
\begin{equation} \label{emt_gen}
T_{\alpha \beta}
= c |g|^{1/2} \, \int p_\alpha p_\beta f \,\frac{dp^1 dp^2 dp^3}{-p_0}.
\end{equation}
Here $|g|$ denotes the modulus of the determinant of the metric, and 
$\Gamma^\alpha_{\beta \gamma}$ are the Christoffel symbols
induced by the metric.
We note that the characteristic system of the Vlasov equation
(\ref{vlasov_gen}) are the geodesic equations written as a first
order system on the mass shell $PM$ which is invariant under 
the geodesic flow.
For more background on the Einstein-Vlasov equation we refer to
\cite{And05}.

The system possesses a large collection of static, spherically
symmetric solutions obtained for example in \cite{Rein94,RR93,RR00}.
The stability properties of these steady states have essentially not
yet been investigated in the mathematics literature;
numerical investigations of this question were reported on
in \cite{AnRe1}, and in \cite{Wo} the author employed variational
methods for constructing steady states as minimizers of
certain energy-Casimir type functionals. 
Examples from the astrophysics literature  
where this stability issue is discussed 
include \cite{IT68,KM,ZeNo,ZePo}.

In contrast,
for the Vlasov-Poisson system, which arises from the
Einstein-Vlasov system in the non-relativistic limit,
considerable mathematical progress on the question of the stability of 
spherically symmetric steady states has been made,
cf.\ \cite{Gu, GuRe01,GuRe,HaRe,LeMeRa2,LeMeRa1,LeMeRa3} 
and the references in the review articles \cite{Mou,Rein07}.
For isotropic steady states
where the particle distribution $f_0$ depends only on the local or 
particle energy the basic stability condition is that
this dependence is strictly decreasing on the support of $f_0$.
In addition to the conserved energy or Hamiltonian
the transport structure of the Vlasov equation
gives rise to a continuum of conserved quantities, the so-called Casimir
functionals. Using the latter the dynamics of the system
can in a natural way be restricted to a leaf $\mathcal{S}_{f_0}$
of perturbations $g:\R^6\to\R$ which have the same level sets as 
the steady state $f_0$ under investigation. On this leaf it is then 
possible to establish a coercivity estimate for the
second variation of the Hamiltonian about $f_0$. 
Such an estimate sometimes goes by the name of Antonov's 
stability bound. 
Under the assumption of spherical symmetry this estimate was used 
in \cite{GuRe,LeMeRa1} to obtain
non-linear stability against spherically symmetric perturbations 
for a large class of steady states $f_0$. 
The case of general perturbations was finally completed in \cite{LeMeRa3}.

For the present case of the Einstein-Vlasov system we intend
to employ an analogous approach. 
However, the numerical investigations in \cite{AnRe1, ZePo} indicate
an essential difference in the stability behavior between the
Vlasov-Poisson and the Einstein-Vlasov systems. For a given
microscopic equation of state, i.e., a given dependence of $f_0$
on the particle energy, there typically exists a one-parameter
family of corresponding steady states. For the Vlasov-Poisson system
all the members of one such family show the same stability behavior,
but this is not so for the Einstein-Vlasov system. 
This remarkable phenomenon was first conjectured and numerically observed 
in the physics literature,
most notably in the work of Ze'ldovitch \cite{ZeNo,ZePo}.
Here, a family of steady states with the same microscopic equation 
of state is parametrized by the central redshift
which is a measure of how close to Newtonian the steady state is.
Provided that the microscopic equation of state 
is strictly decreasing, the steady states in such a family are
stable when they are close to the Newtonian regime,
i.e., when the central redshift is small, but they become unstable
as this parameter increases beyond a certain threshold.
For more precise statements of this behavior and the role
of the so-called fractional binding energy
in this context we refer to
\cite{AnRe1} and the original literature \cite{ZeNo,ZePo}.

Any stability analysis for the Einstein-Vlasov system must be
able to reflect and deal with the above essential differences 
to the Newtonian case,
and the present one does. We prove
linear stability of a suitably defined one-parameter family 
of steady states with small central redshift,
cf.~Theorem~\ref{th:main}. 
The main tool for the result
is a quantified coercivity estimate 
for the second variation of the energy, 
analogous to the Antonov coercivity bound in the Newtonian case,
cf.~Theorem~\ref{th:coercivity}.
This is a non-linear estimate that crucially depends 
both on the symplectic structure of the Einstein-Vlasov system
and on properties of the Einstein field equations satisfied by the 
steady states. 

The paper proceeds as follows.
In the next section we formulate
the system in coordinates which are suitable for the stability
analysis and explicitly introduce $\gamma=1/c^2$ as a parameter. 
We state the form of the ADM mass (or energy) and the Casimir
functionals in these coordinates, and compute the quadratic
term $D^2 {\cal H}_C (f_0)$
in the expansion about the steady state $f_0$
of a suitable energy-Casimir functional  
which is chosen such that the linear part in
the expansion vanishes. 
In Section~\ref{sect_dynac}
we identify a class of linearly dynamically
accessible perturbations which form the natural tangent
space of the leaf $\mathcal{S}_{f_0}$ through $f_0$
which consists of all states which
preserve all the Casimir constraints. The main result of our paper
is shown in Section~\ref{sect_posdef}: 
We prove that the second variation of the energy-Casimir 
is positive definite along the linearly dynamically accessible 
perturbations,
provided that the parameter $\gamma =1/c^2$ is small enough.
As in the case of the Vlasov-Poisson system
this positivity result
should play an important role in a future, fully non-linear 
stability analysis.
In the present paper we restrict ourselves to drawing a conclusion
on linearized stability.
In order to derive such a result
we analyze in Section~\ref{sect_lin} the linearized 
Einstein-Vlasov system. We provide a suitable
existence and uniqueness theory for the corresponding initial value
problem and show that the class of
linearly dynamically accessible perturbations
is invariant under the flow of the linearized system
and that this flow preserves the
quantity $D^2 {\cal H}_C (f_0)$.
In Section~\ref{sect_linstab}
we finally state and prove our result on linearized stability
where we also deal with the fact that for our coercivity
estimate the quantity $\gamma =1/c^2$, which in
a given set of units is a specific constant, has to be chosen 
sufficiently small.  
This is done by exploiting a scaling symmetry of the problem
which relates the Einstein-Vlasov system with 
unit speed of light $c=1$ to the problem where the speed of 
light assumes a prescribed value $c$.
In this analysis the smallness requirement for $\gamma =1/c^2$
translates into the requirement that the steady states under investigation
are close to Newtonian.
As explained above, such a sensitivity of the stability properties
to being close to Newtonian or very relativistic is to be expected.

Our analysis is restricted
to spherically symmetric steady states $f_0$
and their stability against spherically symmetric perturbations.
In particular the latter restriction is undesirable from a
physics point of view. To remove it remains an important
and highly challenging open problem;
the existence of axially symmetric steady states was shown
recently in \cite{AKR}.

To conclude this introduction we mention that 
the global existence result for the spherically symmetric
Einstein-Vlasov system with small initial data
\cite{RR92a} can be considered as a stability result for the vacuum
solution, but the techniques required for the stability
analysis of non-trivial steady states are completely
different from such small data results.
Moreover, the question of weak cosmic censorship and black hole formation
for the asymptotically flat Einstein-Vlasov system
is studied in~\cite{DR} and the latter question is addressed by different methods
in~\cite{AKR_bh}.
A linearized stability analysis for 
the Vlasov-Poisson system is much easier than for the present case
and was performed in \cite{BMR}. 
Concerning the Hamiltonian approach for the Einstein-Vlasov system 
the reference
\cite{KM} has been a most valuable  source of inspiration.
\section{The spherically symmetric Einstein-Vlasov system
         and energy-Casimir functionals} 
\label{sect_ssev}
\setcounter{equation}{0}
We consider a spherically symmetric and asymptotically flat
spacetime. 
In Schwarzschild coordinates the metric takes the form
\begin{equation}\label{metric}
ds^2=-c^2 e^{2\mu(t,r)}dt^2 + e^{2\lambda(t,r)}dr^2+
r^2(d\theta^2+\sin^2\theta\,d\varphi^2).
\end{equation}
Here $t\in\R$ is the time coordinate, $r\in [0, \infty[$ is the
area radius, i.e., $4 \pi r^2$ is the area of the orbit of the
symmetry group $\mathrm{SO}(3)$ labeled by $r$, and the angles
$\theta\in[0, \pi]$ and $\varphi\in[0, 2\pi]$ parametrize these
orbits; $c$ denotes the speed of light.
The spacetime is required to be asymptotically flat
with a regular center which corresponds to the boundary conditions
\[
\lim_{r\to \infty} \lambda(t,r)= \lim_{r\to \infty} \mu(t,r) = 0 = \lambda(t,0).
\] 
In order to formulate the Einstein-Vlasov system we write
$x=(x^a)=r(\sin \theta \cos\phi, \sin \theta \sin \phi, \cos\theta)$.
Let $p^\alpha$ denote the canonical momentum coordinates
corresponding to the spacetime coordinates $(x^\alpha) = (t,x^1,x^2,x^3)$
and define
\[
v^a = p^a + (e^\lambda - 1) \frac{x\cdot p}{r} \frac{x^a}{r},
\ \mbox{where}\ x\cdot p = \delta_{ab}x^a p^b.
\]
This has the advantage that
by the mass shell condition
\[
p_0 = - e^\mu \sqrt{1+\gamma |v|^2},\ \mbox{where}\ |v|^2 = \delta_{ab}v^a v^b
\ \mbox{and}\ \gamma :=\frac{1}{c^2}.
\]
We introduce the abbreviation
\begin{equation}\label{sqrtdef}
\p :=  \sqrt{1+\gamma |v|^2}.
\end{equation}
In the canonical momentum variables the metric would
also appear under the square root sign. The  Einstein-Vlasov system
now takes the following form:
\begin{equation} \label{vlasov}
\partial_t f + e^{\mu - \lambda}\frac{v}{\p }\cdot \partial_x f -
\left( \dot \lambda \frac{x\cdot v}{r} + e^{\mu - \lambda} \frac{1}{\gamma} \mu'
\p  \right) \frac{x}{r} \cdot \partial_v f =0,\
\end{equation}
\begin{eqnarray}
e^{-2\lambda} (2 r \lambda' -1) +1 
&=& 
8\pi \gamma r^2 \rho ,\label{eelambda} \\
e^{-2\lambda} (2 r \mu' +1) -1 
&=& 
8\pi \gamma^2 r^2 p,\label{eemu}\\
- e^{-\mu - \lambda} \dot \lambda 
&=&
4 \pi \gamma r \jmath \label{eelambdad},
\end{eqnarray}
\begin{equation}
e^{- 2 \lambda} \left(\mu'' + (\mu' - \lambda')(\mu' + \frac{1}{r})\right)
- e^{-2\mu}\left(\ddot \lambda + \dot\lambda (\dot \lambda - \dot \mu)\right)
= 4 \pi \gamma^2 q,\label{ee2ndo} 
\end{equation}
where
\begin{eqnarray}
\rho(t,x) 
&=& 
\int  f(t,x,v)\p \,dv ,\label{rhodef}\\
p(t,x) 
&=& 
\int  f(t,x,v)\left(\frac{x\cdot v}{r}\right)^2
\frac{dv}{\p },\label{pdef}\\
\jmath(t,x) 
&=& 
\int  f(t,x,v) \frac{x\cdot v}{r} dv,\label{jdef}\\
q(t,x) 
&=& 
\int  f(t,x,v) \left|\frac{x\times v}{r}\right|^2
\frac{dv}{\p}. \label{qdef}
\end{eqnarray}
Here $\partial_xf$ and $\partial_vf$ denote the gradients of $f$ with respect to
the $x$- and $v$ variable respectively, and
$\dot{\phantom\lambda}$ and ${\phantom \lambda}'$
denote partial derivatives with respect to $t$ or $r$ respectively.
Unless explicitly noted otherwise integrals extend
over $\R^3$.
Spherical symmetry means that
\[
f(t,x,v) = f(t,A x,A v),\ x, v \in \R^3,\ A \in {\rm SO}\,(3),
\]
i.e., $f$ is invariant under the canonical action of ${\rm SO}\, (3)$
on the mass shell. As a consequence the spatial densities defined
in (\ref{rhodef})--(\ref{qdef}) are actually functions of $t$ and
$r=|x|$. We refer to \cite{Rein95} for details on the
derivation of this form of the equations. It has been used both
in investigations of global existence for small data
\cite{RR92a} and of the formation of black holes
\cite{AKR_bh}. The former investigation also
provides a local existence and uniqueness theorem
for the spherically symmetric Einstein-Vlasov system
for smooth, compactly supported and non-negative initial data
\begin{equation} \label{inidata}
f_{|t=0} = \fn \in C^1_c (\R^6),\ \fn \geq 0,
\end{equation}
which are compatible
with (\ref{eelambda}) in the sense explained shortly.

In a stability analysis the following
aspect of the choice of coordinates should be noted:
In order to compare unperturbed and perturbed quantities 
one must define some identification of points in the unperturbed
with points in the perturbed spacetime where the various quantities
are to be evaluated. A priori, no natural such identification exists,
and for the Einstein-Vlasov system this question must be faced
not only for points on spacetimes but for points on  their tangent
bundles or corresponding mass shells respectively.
Throughout our analysis we identify points which have the same
coordinates in the above set-up. Besides the fact that this choice
works it is from a physics point of view
motivated in \cite{IT68}. 

For what follows it is important to note that
given a spherically symmetric state $f\in C^1_c (\R^6)$
we can explicitly and uniquely solve the field equation (\ref{eelambda}) for
$\lambda$ under the boundary condition $\lambda(0)=0$,
\begin{equation}\label{lambdadef}
e^{-2\lambda(r)} = 1 - \gamma \frac{2  m (r)}{r},
\end{equation}
where
\[
m(r) = 4 \pi \int_0^r s^2 \rho(s)\, ds 
= \int_{|x|\leq r}\int \p f(x,v)\,dv\,dx;
\]
we will occasionally write $\lambda_f$, $\rho_f$, $m_f$ to emphasize
that these quantities are determined by $f$. Clearly, for
(\ref{lambdadef}) to define $\lambda$ on all of $[0,\infty[$
we have to restrict the set of admissible states.
We call a state $f\in C^1_c (\R^6)$ {\em admissible} iff
it is non-negative, spherically symmetric, and
\[
\gamma \frac{2 m_f (r)}{r} < 1,\ r\geq 0;
\] 
the initial data posed in (\ref{inidata}) must be admissible 
in this sense.

We now define the ADM mass and the Casimir functionals 
as functionals on the set of admissible states:
\begin{eqnarray}
\mathcal{H}(f) \label{hdef}
&=&
\iint\p f\,dv\,dx,\\
\mathcal{C}(f)
&=&
\iint e^{\lambda_f}\chi(f)\,dv\,dx, \label{casidef}
\end{eqnarray}
where $\chi \in C^1(\R)$ with $\chi(0)=0$. These quantities are conserved
along solutions launched by admissible initial data:
\[
\mathcal{H}(f(t))=\mathcal{H}(\fn\,),\ \mathcal{C}(f(t))=\mathcal{C}(\fn\,)
\]
as long as the solution exists; for functions $f=f(t,x,v)$ we
write $f(t):= f(t,\cdot,\cdot)$, and the analogous notational
convention applies to functions of $t$ and $x$ or $t$ and $r$. 

We need to recall some facts about
steady states of the system (\ref{vlasov})--(\ref{jdef}),
a more detailed discussion including their dependence on
$\gamma$ is given at the beginning of Section~\ref{sect_posdef}.
For $\mu=\mu_0(r)$ time-independent and given we 
define the local or particle energy
\begin{equation} \label{edef}
E = E(x,v)= e^{\mu_0(r)} \p = e^{\mu_0(r)} \sqrt{1+\gamma |v|^2}.
\end{equation}
The ansatz 
\begin{equation} \label{ssansatz}
f_0(x,v)) = \phi (E(x,v))
\end{equation}
with some suitable microscopic equation of state $\phi=\phi(E)$
then satisfies the time-independent Vlasov equation and
reduces the problem of finding stationary solutions to
solving the two field equations (\ref{eelambda}), (\ref{eemu})
where the source terms $\rho_0$ and $p_0$ are now functionals 
of $\mu=\mu_0$.
A necessary condition for obtaining a steady state
with compact support and finite ADM mass is that
$\phi(E) = 0$ for $E>E_0$ where $E_0>0$ is some cut-off energy.
We assume that $\phi \in C^2(]-\infty,E_0[)\cap C(\R)$ 
is strictly decreasing on $]-\infty,E_0]$
and such that a corresponding compactly supported steady state 
$(f_0,\lambda_0,\mu_0)$ with induced spatial densities
$\rho_0, p_0$ and finite ADM mass exists. 
Examples of such functions are provided in  
\cite{Rein94,RR93,RR00}.

We now discuss on a formal level the energy-Casimir
approach towards stability for the Einstein-Vlasov system. 
We consider an energy-Casimir functional
\begin{equation} \label{ecdef}
\mathcal{H}_C (f) := \mathcal{H} (f) + \mathcal{C} (f),
\end{equation}
and we expand about some given steady state $f_0$
of the above form:
\begin{equation} \label{ecexpand}
\mathcal{H}_C (f_0+\delta f) = 
\mathcal{H}_C (f_0)+D\mathcal{H}_C(f_0)(\delta f)
+ D^2\mathcal{H}_C(f_0)(\delta f,\delta f) + \mathrm{O}((\delta f)^3).
\end{equation}
A rather lengthy and non-trivial formal computation reveals the following.
If the function $\chi$ which generates the Casimir functional 
(\ref{casidef}) is chosen such that
\[
\chi'(f_0) = \chi' (\phi(E)) = - E,\ \mbox{i.e.}\ \chi' = - \phi^{-1},
\]
then
\[
D\mathcal{H}_C(f_0)(\delta f) =0
\]
and
\begin{eqnarray}
D^2\mathcal{H}_C(f_0)(\delta f,\delta f)
&=&
\frac{1}{2}\iint \frac{e^{\lambda_0}}{|\phi'(E)|}(\delta f)^2\,dv\,dx
\nonumber \\
&&
{}- \frac{1}{2\gamma}\int_0^\infty e^{\mu_0 - \lambda_0}
\left(2 r \mu_0' +1\right)\, (\delta\lambda)^2\,dr.\label{qecdef}
\end{eqnarray}
Here $\delta\lambda$ should be expressed in terms of $\delta f$ through the
variation of (\ref{lambdadef}), cf.\ (\ref{varlambdadef}) below,
and it should be observed that on the support of the steady
state $\phi$ is strictly decreasing and in particular one-to-one.
Since $\phi'(E)=0$ outside the compact support of the steady state
$f_0$, the first integral makes sense only for perturbations
$\delta f$ which are supported in the support of the steady state.
This is automatically the case for the class of linearly dynamically
accessible states defined in Section~\ref{sect_dynac}.
Just as in the case of the Vlasov-Poisson system the central difficulty 
in the stability analysis arises from the fact that the two terms
in (\ref{qecdef}) are of opposite sign. For the Vlasov-Poisson system
is has been shown in \cite{KS} that $D^2 {\cal H}_c(f_0)$ is
positive definite on a suitably defined class of linearly dynamically
accessible states, and this fact has played a central role in
the non-linear stability analysis in \cite{GuRe,HaRe}.
In Section~\ref{sect_posdef} we prove an analogous positivity
result for the case of the Einstein-Vlasov system, and we believe
that this should be a useful step towards a non-linear stability
result. In this context the precise relation of $D^2 {\cal H}_c(f_0)$
to the energy-Casimir functional (\ref{ecdef}) is important which
is why we have gone into the issue of the expansion (\ref{ecexpand}).
However, in the present paper we only exploit $D^2 {\cal H}_c(f_0)$
and its properties to obtain a linear stability result.
For this the relation (\ref{ecexpand}) is in principle irrelevant,
the important issue being that $D^2 {\cal H}_c(f_0)$ is a conserved
quantity along solutions of the linearized Einstein-Vlasov system,
linearized about the steady state $f_0$. This is shown is 
Section~\ref{sect_lin}. The consequences of our results for non-linear
stability are under investigation, and in this context the
expansion (\ref{ecexpand}) will have to be made mathematically precise.

To conclude this section we recall the usual Poisson bracket
\begin{equation} \label{pbdef}
\{f,g\}:=\partial_x f \cdot\partial_v g-\partial_v f \cdot\partial_x g 
\end{equation}
for two continuously differentiable functions $f$ and $g$ of
$x,v \in \R^3$. We shall repeatedly use the product rule
\begin{equation} \label{productrule}
\{f,g h\}= \{f,g\} h + \{f,h\} g 
\end{equation}
and the integration by parts formula
\begin{equation} \label{intbyparts}
\iint\{f,g\}\,dv\,dx = 0
\end{equation}
which together with the product rule implies that
\begin{equation} \label{intbypartsimpl}
\iint\{f,g\} h \,dv\,dx = - \iint\{f,h\} g \,dv\,dx.
\end{equation}
Since throughout, our functions are spherically symmetric
it is sometimes convenient to use the following coordinates which are adapted
to this symmetry:
\[
r=|x|,\ w = \frac{x\cdot v}{r},\ L= |x\times v|^2;
\]
a distribution function $f=f(x,v)$ is spherically symmetric iff by
abuse of notation,
\[
f=f(r,w,L).
\]
It is also worthwhile to note that $L$ is conserved along
the characteristics of the Vlasov equation (\ref{vlasov}),
i.e., due to spherical symmetry angular momentum is conserved
along particle trajectories. 
If $f$ and $g$ are spherically symmetric 
and written in terms of $r,w,L$, then
\begin{equation} \label{pbrwL}
\{f,g\}=\partial_r f \, \partial_w g-\partial_w f \, \partial_r g. 
\end{equation}
Notice further that
\[
\frac{1}{\gamma} \{f,E\} = 
e^{\mu_0} \frac{v}{\p }\cdot \partial_x f - 
\frac{1}{\gamma} e^{\mu_0} \mu_0' \p 
\frac{x}{r}\cdot \partial_v f,
\]
which should be compared with the Vlasov equation (\ref{vlasov}).
\section{Dynamically accessible states}
\label{sect_dynac}
\setcounter{equation}{0}
Let $(f_0,\lambda_0,\mu_0)$ with (\ref{ssansatz}) be a fixed 
stationary solution of the spherically symmetric Einstein-Vlasov 
system whose stability we want to investigate.
We call an admissible state $f$ 
{\em non-linearly dynamically accessible from $f_0$} iff
for all $\chi \in C^1(\R)$ with $\chi(0)=0$,
\begin{equation}\label{nldynac}
\mathcal{C}(f)=\mathcal{C}(f_0).
\end{equation}
This property is preserved by the flow of
the Einstein-Vlasov system. The aim of the present section
is to develop a suitable concept of linearly dynamically accessible
states on which the second variation of $\mathcal{H}_C$ at $f_0$
is positive definite while the first one vanishes.
Moreover, the set of these linearly dynamically accessible states should be
invariant under the flow of the linearized Einstein-Vlasov system.
Formally, taking the first variation in (\ref{nldynac}), a suitable 
definition for $\delta f$ to be linearly dynamically accessible should
be that
\begin{equation}\label{ldynac}
D\mathcal{C}(f_0)(\delta f)
=\iint e^{\lambda_0} \left(\chi'(f_0)\delta f
+\chi(f_0)\delta\lambda\right) dv\,dx = 0
\end{equation}
for all $\chi \in C^1(\R)$ with $\chi(0)=0$, where
\begin{equation} \label{varlambdadef}
\delta\lambda = 
\gamma e^{2 \lambda_0}\frac{4 \pi}{r} \int_0^r s^2 \delta\rho (s)\,ds
\end{equation}
and
\begin{equation} \label{varrhodef}
\delta\rho(r) = \delta\rho(x) = \int\p  \delta f (x,v)\, dv.
\end{equation}
Condition (\ref{ldynac}) is from an analysis point of view not
practical to work and prove estimates with. In order to obtain 
a more explicit condition on $\delta f$ we first note a simple fact.

\begin{lemma}\label{intpart}
Under the assumptions made above,
\[
\int  \chi (f_0)\, dv = -\gamma
e^{\mu_0}\int \chi'(f_0)\,\phi'(E)\frac{w^2}{\p }dv,
\]
and hence
\[
D\mathcal{C}(f_0)(\delta f)
=\iint e^{\lambda_0}\chi'(f_0)\left[\delta f  
- \gamma \,e^{\mu_0}\delta\lambda \, \phi'(E)\frac{w^2}{\p }\right] dv\,dx.
\]
\end{lemma}
{\bf Proof.}
We use the expressions 
$E=e^{\mu_0}\p $ and $L=|x\times v|^2$ as new 
integration variables and observe that the integrand is even in $v$.
Hence
\begin{eqnarray*}
dv
&=&
\frac{2\pi e^{-\mu_0}}{\sqrt{\gamma}r^2}
\frac{E}{\sqrt{E^2-e^{2\mu_0}\left(1+\gamma L/r^2\right)}}\,dE\,dL\\
&=&
\frac{2\pi e^{-\mu_0}}{\sqrt{\gamma}r^2}
\partial_E \sqrt{E^2-e^{2\mu_0}\left(1+\gamma L/r^2\right)}
\,dE\,dL.
\end{eqnarray*}
An integration by parts now gives the result:
\begin{eqnarray*}
&&
\int\chi(f_0)\,dv\\
&&
=\frac{2\pi e^{-\mu_0}}{\sqrt{\gamma}r^2}
\int_{0}^{\infty}\int_{e^{\mu_0}\sqrt{1+\gamma \frac{L}{r^2}}}^\infty\chi(\phi(E))
\partial_E \sqrt{E^2-e^{2\mu_0}\left(1+\gamma \frac{L}{r^2}\right)}\,dE\,dL\\
&&
=
-\frac{2\pi e^{-\mu_0}}{\sqrt{\gamma}r^2}
\int_{0}^{\infty}\int_{e^{\mu_0}\sqrt{1+\gamma \frac{L}{r^2}}}^\infty
\chi'(\phi(E))\phi'(E)
\sqrt{E^2-e^{2\mu_0}\left(1+\gamma \frac{L}{r^2}\right)}\,dE\,dL\\
&&
=
-\int 
\chi'(\phi(E))\phi'(E)
\frac{E^2-e^{2\mu_0}(1+\gamma L/r^2)}{E} dv\\
&&
=
-\gamma e^{\mu_0}\int \chi'(f_0)\phi'(E) \frac{w^2}{\p } dv .
\end{eqnarray*}
\prfe
Using this result we find that a variation $\delta f$ satisfies
the condition (\ref{ldynac}), if
\begin{equation} \label{ldynacexpl}
e^{\lambda_0} \delta f -\gamma e^{\mu_0+\lambda_0}\delta\lambda\, 
\phi'(E)\frac{w^2}{\p }=\{h,f_0\}
\end{equation}
for some spherically symmetric generating function $h\in C^2(\R^6)$.
This is due to the fact that
\[
\iint\chi'(f_0)\,\{h,f_0\}\,dv\,dx =0
\]
for any such generating function $h$.
Since $\delta\lambda$ appears in the second term, (\ref{ldynacexpl})
is still not suitable as a definition for $\delta f$, but we have the 
following result:

\begin{proposition} \label{ldynacprop}
Let
\begin{equation} \label{ldynacdef}
\delta f :=
e^{-\lambda_0}\{h,f_0\}
+4\pi \gamma^3 r e^{2\mu_0+\lambda_0}\phi'(E)
\frac{w^2}{\p }\int \phi'(E(x,\tilde v))\,h(x,\tilde v)\,\tilde w\,d\tilde v
\end{equation}
for some spherically symmetric generating function $h\in C^2(\R^6)$.
If the corresponding variation of $\lambda$ is defined by
(\ref{varlambdadef}), then
\begin{equation}\label{ldynaclambda}
\delta\lambda=4\pi r \gamma^2 e^{\mu_0+\lambda_0}
\int \phi'(E)\,h(x,v)\,w\,dv.
\end{equation}
Hence $\delta f$ satisfies both (\ref{ldynacexpl}) and
(\ref{ldynac}).
States of the form (\ref{ldynacdef}) are called
{\em linearly dynamically accessible from $f_0$}.
\end{proposition}
For the proof the following auxiliary result is needed
which will also be used elsewhere.
\begin{lemma}\label{intphipr}
The following identity holds:
\[
\int \phi'(E) w^2 dv
=
- \frac{e^{-\mu_0}}{\gamma}\left(\gamma p_0 + \rho_0\right)
=
-\frac{e^{-2\lambda_0-\mu_0}}{4\pi \gamma^2 r}
\left(\lambda_0'+\mu_0'\right).
\]
\end{lemma}
{\bf Proof.}
We note that
\[
\partial_v \phi(E) = \phi'(E) e^{\mu_0}\frac{\gamma v}{\p }
\]
and hence
\[
\frac{x}{r}\cdot \partial_v \phi(E) = 
\phi'(E) e^{\mu_0}\frac{\gamma w}{\p }.
\]
This implies that
\begin{eqnarray*}
\int \phi'(E) w^2 dv
&=&
\frac{e^{-\mu_0}}{\gamma} \int \frac{x}{r}\cdot \partial_v \phi(E)
\p  w\, dv\\
&=&
- \frac{e^{-\mu_0}}{\gamma} \int \phi(E)
\left(\frac{\gamma w^2}{\p }+
\p \right)  dv\\
&=&
- \frac{e^{-\mu_0}}{\gamma}\left(\gamma p_0 + \rho_0\right)
=-\frac{e^{-2\lambda_0-\mu_0}}{4\pi \gamma^2 r}
\left(\lambda_0'+\mu_0'\right);
\end{eqnarray*}
the last equality is due to the field equations
(\ref{eelambda}), (\ref{eemu}) for the steady state. \prfe
{\bf Proof of Proposition~\ref{ldynacprop}.}
The variation of $\rho$ induced by
$\delta f$ takes the form
\[
\delta\rho
=
e^{-\lambda_0}\int\p \{h,f_0\} dv
{}+ 4\pi r \gamma^3 e^{\lambda_0+2\mu_0} \int\phi'(E) h w\, dv
\int\phi'(E) w^2 dv.
\]
Hence Lemma~\ref{intphipr} implies that
\[
\delta\rho =
e^{-\lambda_0}\int\p \{h,f_0\} dv
-\gamma e^{\mu_0-\lambda_0} \left(\lambda_0'+\mu_0'\right) \int\phi'(E) h w\, dv.
\]
We consider the first term on the right hand side and find that
\begin{eqnarray*}
&&
\int\p \{h,f_0\} dv\\
&&
=
\int\p 
\left(\partial_x h \cdot\frac{e^{\mu_0}\gamma v}{\p }
-\partial_v h\cdot\frac{x}{r}\mu_0'e^{\mu_0} \p \right)\, 
\phi'(E)\,dv\\
&&
=
\gamma e^{\mu_0}\int \partial_x h \cdot v \phi'(E)\,dv
-\mu_0'e^{\mu_0}\int \partial_v h\cdot\frac{x}{r}
(1+\gamma |v|^2)\phi'(E)\,dv\\
&&
=
\gamma e^{\mu_0}\int\biggl[ \partial_x h \cdot v \phi'(E)
+2 \mu_0'w h \phi'(E)
+ e^{\mu_0}\mu_0'w h \phi''(E)\p \biggr]\, dv.
\end{eqnarray*}
Hence
\begin{eqnarray*}
\delta\lambda
&=&
e^{2\lambda_0}\frac{\gamma^2}{r}\int_{|x|\leq r}\int
e^{\mu_0-\lambda_0} \biggl[ \partial_x h \cdot v \phi'(E)
+ 2 \mu_0'w h \phi'(E)\\
&&
\qquad\qquad\qquad\qquad 
+ e^{\mu_0}\mu_0'w h \phi''(E)\p 
- \left(\lambda_0'+\mu_0'\right) \phi'(E) h w \biggr] dv\\
&=&
e^{2\lambda_0}\frac{\gamma^2}{r} 4 \pi r^2 e^{\mu_0-\lambda_0}
\int w h \phi'(E)\, dv\\
&&
{}+e^{2\lambda_0}\frac{\gamma^2}{r}
\int_{|x|\leq r}\int e^{\mu_0-\lambda_0} \biggl[
-(\mu_0'-\lambda_0')w \phi'(E) - 
\phi''(E) e^{\mu_0}\mu_0'w \p \\
&&
\qquad\qquad\qquad\qquad\qquad\qquad
{}+ 2 \mu_0'w \phi'(E) + \mu_0'e^{\mu_0} w \p \phi''(E)\\
&&
\qquad\qquad\qquad\qquad\qquad\qquad
{}-\left(\lambda_0'+\mu_0'\right) \phi'(E) w \biggr] h\, dv\, dx,
\end{eqnarray*}
and since the term in brackets vanishes the claim is proven.
\prfe

\noindent
{\bf Remark.} For linearly dynamically accessible states as defined
in Proposition~\ref{ldynacprop},
\begin{eqnarray*}
\delta f 
&=&
\phi'(E)\,\left(e^{-\lambda_0}\{h,E\}
+4\pi \gamma^3 r e^{2\mu_0+\lambda_0}
\frac{w^2}{\p }\int \phi'(E)\,h(x,v)\,w\,dv\right)\\
&=&
\phi'(E)\,\left(e^{-\lambda_0}\{h,E\}
+ \gamma r e^{\mu_0}\delta\lambda \,\frac{w^2}{\p }\right),
\end{eqnarray*}
in particular, $\delta f$ vanishes outside the support of
$f_0$ and the integrals in (\ref{qecdef}) are well defined.
\section{The coercivity estimate} 
\label{sect_posdef}
\setcounter{equation}{0}

The central step in our stability analysis is a coercivity estimate 
for the second variation of the energy. We recall the definition
of this expression which is also referred to
as the free energy:
\begin{eqnarray}\label{secvar}
\mathcal{A}(\delta f)
&:=&
D^2\mathcal{H}_C(f_0)(\delta f,\delta f) \\
&=&
\frac{1}{2}\iint \frac{e^{\lambda_0}}{|\phi'(E)|}(\delta f)^2\,dv\,dx
- \frac{1}{2\gamma}\int_0^\infty e^{\mu_0 - \lambda_0}
\left(2 r \mu_0' +1\right)\, (\delta\lambda)^2\,dr.\nonumber
\end{eqnarray}
The estimate will hold provided
$\gamma$ is sufficiently small. We need to make precise
the assumptions on the steady states under consideration and
need to discuss their behavior for $\gamma \to 0$,
i.e., their relation to those of the Vlasov-Poisson system 
\begin{equation}
\partial_{t}f+v\cdot \partial_{x}f-\partial_{x}U\cdot \partial_{v}f =0,
\label{N-vlasov}
\end{equation}
\begin{equation}
\Delta U = 4\pi \rho,\ \lim_{|x| \to \infty} U(t,x) = 0, \label{poisson}
\end{equation}
\begin{equation}
\rho(t,x) = \int f(t,x,v)\,dv. \label{N-rhodef} 
\end{equation}
Here $U=U(t,x)$ is the gravitational potential of the ensemble, 
and $\rho=\rho(t,x)$ is its spatial mass density. 
The Newtonian particle energy is given by
\begin{equation} \label{N-partendef}
E=E(x,v) = \frac{1}{2} |v|^2 + U(x),
\end{equation}
and under suitable assumptions on $\Phi$ the ansatz 
\be\label{eq:nansatz}
f_0(x,v)=\Phi\left(\frac{1}{2}|v|^2+U(x)\right) 
\ee
leads to spherically symmetric steady states.
 
\noindent
{\bf Assumption on $\Phi$.} Let
$\Phi \in C(\R)\cap C^2(]-\infty,0[)$ with 
\[
\Phi = 0\ \mbox{on}\ [0,\infty[ \ \mbox{and}\ 
\Phi'< 0 \ \mbox{on}\ ]-\infty,0[
\]
be such that for a parameter $\open{\nu}<0$
the semilinear Poisson equation
\begin{equation} \label{semilinpoisson}
\frac{1}{r^2} (r^2 U'(r))' = 
4\pi \int_{U(r)}^\infty \Phi(E)\sqrt{2(E-U(r))}\, dE
\end{equation}
has a unique solution $U\in C^2([0,\infty[)$ with
the central value $U(0)=\open{\nu}$ and the property
that $U(R_0)=0$ for some radius $R_0>0$.

The ansatz~(\ref{eq:nansatz}) 
necessarily leads to a spherically symmetric steady state where
$U=U(r),\ r=|x|$, and
reduces the static Vlasov-Poisson system to the equation 
(\ref{semilinpoisson}); the corresponding steady state is
supported in space in the ball of radius $R_0$ about the origin.
We also remark that the boundary condition for the potential
at spatial infinity has been replaced by the condition that
the potential vanishes at the boundary of the support of the
steady state. This is technically advantageous, and the original
boundary condition can be restored by a suitable shift of the potential
and a corresponding cut-off energy $E_0<0$. 
A well known example where our assumptions hold are
the polytropic steady states given by
\[
\Phi(E)=\left\{\begin{array}{ccl}(-E)^k&,&\ E<0,\\
                                 0&,&\ E\geq 0
                \end{array} \right.
\]
with $k\in ]0,7/2[$; more examples can be found in \cite{RR00}.

For the Einstein-Vlasov system the particle energy is given by
\[
E = E(x,v)= e^{\mu_0(r)} \p = e^{\mu_0(r)} \sqrt{1+\gamma |v|^2}.
\]
We fix a function $\Phi$ as above and write $\mu_0 = \gamma \nu_0$.
By the ansatz
\[
f_0(x,v) = 
\Phi\left(
\frac{1}{\gamma} \sqrt{1+\gamma |v|^2}e^{\gamma \nu_0}-\frac{1}{\gamma}
\right) = \Phi\left(
\frac{E-1}{\gamma}\right)
\] 
the static Einstein-Vlasov system is reduced to a single equation
for $\nu_0$, namely
\begin{equation} \label{ngl}
\nu_0'(r) = 
\frac{4\pi}{1- \frac{8\pi}{r} \gamma \int_0^r s^2 g_{\gamma} (\nu_0(s)) ds}
\left(\gamma r h_{\gamma} (\nu_0(r)) + \frac{1}{r^2} 
\int_0^r s^2 g_{\gamma} (\nu_0(s))\, ds \right) ,
\end{equation}
where $g_{\gamma}$ and $h_{\gamma}$ are smooth functions determined
by $\Phi$ which are such that
$\rho_0(r)=g_{\gamma} (\nu_0(r))$ and $p_0(r)=h_{\gamma} (\nu_0(r))$.
The important point is that 
\begin{equation} \label{glimit}
g_{\gamma} (\nu_0(r))\to 
\int_{\nu_0(r)}^\infty \Phi(E)\sqrt{2(E-\nu_0(r))}\, dE
\ \mbox{as}\ \gamma \to 0.
\end{equation}
It should be noticed that the ansatz (\ref{ssansatz}) has been
adapted to yield the proper Newtonian limit, and we have the relation 
$\phi(\cdot)=\Phi(\frac{\cdot}{\gamma}-\frac{1}{\gamma})$.
For the details of these arguments we have to refer to \cite{RR93};
here we collect only the information which we need for the stability
analysis.
\begin{proposition}\label{ssprop}
There exist constants $\gamma_0 >0$ and $C>0$ such that
for $0 < \gamma \leq \gamma_0$ the equation (\ref{ngl}) has a unique
solution $\nu_0 \in C^2([0,\infty[)$ with $\nu_0(0)=\open{\nu}$.
The resulting steady state $(f_0,\lambda_0,\mu_0=\gamma\nu_0)$ 
satisfies the following estimates:
\[
|x| + |v| \leq C\ \mbox{and}\ \nu_0' > \frac{1}{C}
\ \mbox{on}\ \mathrm{supp} f_0,
\]
and
\[
||\rho_0||_\infty,\ ||p_0||_\infty,\ ||\lambda_0||_\infty,\
||\nu_0||_\infty,\ ||\nu_0'||_\infty \leq C .
\]
\end{proposition}
{\bf Sketch of the proof.} By the assumption on $\Phi$
the Newtonian potential $U$ has a zero at some radius $R_0$,
and since $U$  is strictly increasing it is strictly positive
for $r>R_0$. The structure of the right hand side of (\ref{ngl})
and in particular the limiting behavior (\ref{glimit}) implies
that $\nu_0$ converges to $U$ uniformly on bounded intervals
as $\gamma \to 0$, in particular, $\nu_0$ must also have a zero
at a radius close to $R_0$ for $\gamma$ small, and the rest
follows; for details we refer to  \cite{RR93}. 
\prfe
The content of the following theorem is the coercivity estimate  
for the free energy $\mathcal{A}$ on linearly
dynamically accessible perturbations.
\begin{theorem}\label{th:coercivity}
There exist constants $C^\ast >0$ and $\gamma^\ast>0$ such that
for any $0<\gamma\leq\gamma^\ast$ and any
spherically symmetric function $h\in C^2(\R^6)$ 
which is odd in the $v$-variable the estimate
\[
\mathcal{A}(\delta f)\geq 
C^\ast\iint|\phi'(E)|\,\left((rw)^2 
\left|\left\{E,\frac{h}{rw}\right\}\right|^2 
+ \gamma^2 |h|^2\right) dv\,dx
\]
holds. Here $\delta f$ is the dynamically accessible perturbation 
generated by $h$ according to (\ref{ldynacdef}).
\end{theorem}
Before proving this theorem we collect some auxiliary results. 
\begin{lemma}\label{auxlem}
Let $h\in C^2(\R^6)$ be spherically symmetric.
Then the following estimate holds:
\[
\left(\int |\phi'(E)| wh\,dv\right)^2
\leq
\frac{e^{-2\lambda_0-\mu_0}}{4\pi\gamma^2r}
\left(\lambda_0'+\mu_0'\right)\int|\phi'(E)|h^2 dv.
\]
\end{lemma}
{\bf Proof.}
By the Cauchy-Schwarz inequality,
\[
\left(\int |\phi'(E)| wh\,dv\right)^2
\leq \int|\phi'(E)|w^2\,dv\int|\phi'(E)|h^2 dv,
\]
and applying Lemma~\ref{intphipr} completes the proof.
\prfe
\begin{lemma}\label{identities}
The following identities hold:
\begin{eqnarray*}
\{E,rw\}
&=&
e^{\mu_0}r\mu_0'\p-e^{\mu_0}\p+\frac{e^{\mu_0}}{\p},\\
\{E,\{E,rw\}\}
&=&
-\gamma e^{2\mu_0} w
\left[r \mu_0'' + \mu_0'+ \frac{2\mu_0'}{\p^2}\right].
\end{eqnarray*}
\end{lemma}
{\bf Proof.}
The first identity follows easily from the product rule
(\ref{productrule}) and the definition (\ref{edef}) of $E$,
for the second identity we use the first one and again the product
rule. \prfe

\noindent
{\bf Proof of Theorem~\ref{th:coercivity}.}
Before starting our estimates we mention that
throughout the following proof, $0<\gamma\leq \gamma_0$ as
in Proposition~\ref{ssprop}, and $C$ denotes
a positive constant which may change its value
from line to line, depends on the bounds provided
in Proposition~\ref{ssprop}, and is independent of $\gamma$.

\smallskip\noindent
{\em Step 1: Splitting $\mathcal{A}$.}
Using the formula~(\ref{ldynacdef}) for $\delta f$ and the 
definition~(\ref{secvar}) of $\mathcal{A}$, we
expand the square of $\delta f$ appearing under the integral sign 
and rewrite $\mathcal{A}(\delta f,\delta\lambda)$ in the following form:
\begin{equation}\label{a1a2}
2 \A(\delta f)=
\A_1(\delta f)+\A_2(\delta f),
\end{equation}
where 
\begin{eqnarray}
\A_1(\delta f)
&:=&
\iint e^{-\lambda_0}|\phi'(E)||\{E,h\}|^2 dv\,dx\nonumber\\
&&
{}-\frac{1}{\gamma}\int_0^{\infty}e^{\mu_0-\lambda_0}
(2r\mu_0'+1)(\delta\lambda)^2 dr,\nonumber\\
&=:&
\A_{11}(\delta f) + \A_{12}(\delta f), \label{a1}\\
\A_2(\delta f)
&:=&
-2 \gamma \iint |\phi'(E)|\{E,h\} 
\delta\lambda e^{\mu_0}\frac{w^2}{\p}\,dv\,dx
\nonumber \\
&&
+ \gamma^2\iint|\phi'(E)|e^{2\mu_0+\lambda_0}\frac{w^4}{\p^2}(\delta\lambda)^2
dv\,dx \nonumber\\
&=:&
\A_{21}(\delta f) + \A_{22}(\delta f).
\label{a2}
\end{eqnarray}
The idea behind this splitting is that $\A_1$ will yield the desired
lower bound while $\A_2$ is of higher order in $\gamma$ and can eventually
be controlled by the positive contribution from $\A_1$.

\smallskip\noindent
{\em Step~2: The term $\A_1$.}
Let us define
\[
\eta:=\frac{1}{rw}h.
\]
The function $\eta$ is well-defined for $w=0$ 
since $h$ is odd in $v$. By the product rule (\ref{productrule}),
\begin{equation}\label{decomp1}
\{E,h\}=rw\{E,\eta\}+\eta\{E,rw\},
\end{equation}
and by (\ref{decomp1}),
\[
|\{E,h\}|^2=
(rw)^2|\{E,\eta\}|^2+\{E,\eta^2rw\{E,rw\}\}-\eta^2rw\{E,\{E,rw\}\}.
\]
Inserting this into $\A_{11}$, we arrive at
\begin{eqnarray}
&&
\iint e^{-\lambda_0}|\phi'(E)|\,|\{E,h\}|^2 dv\,dx 
=\iint e^{-\lambda_0}|\phi'(E)|\,(rw)^2|\{E,\eta\}|^2 dv\,dx\nonumber\\
&&
\qquad \quad {} 
+\iint e^{-\lambda_0}|\phi'(E)|\,\{E,\eta^2rw\{E,rw\}\}\, dv\,dx\nonumber\\
&&
\qquad \quad {}
-\iint e^{-\lambda_0}|\phi'(E)|\,\eta^2rw\{E,\{E,rw\}\}\,dv\,dx\nonumber\\
&&
\qquad
=:U+V+W.\label{u+v+w}
\end{eqnarray}
The term $U$ is a positive definite contribution to $\A_1$. 
The terms $V$ and $W$ are more delicate since they will have to be
combined with the negative term $\A_{12}$.
For the term $V$ we use the formula (\ref{intbypartsimpl})
and integrate by parts in $v$ to arrive at
\begin{eqnarray*}
V
&=&
-\iint e^{-\lambda_0}\{f_0,\eta^2rw\{E,rw\}\}\,dv\,dx\\
&=&
\iint f_0\{e^{-\lambda_0},\eta^2rw\{E,rw\}\}\,dv\,dx\\
&=&
\iint f_0\partial_x(e^{-\lambda_0})\cdot\partial_v(\eta^2rw\{E,rw\}\})\,dv\,dx\\
&=&
-\iint \partial_vf_0\cdot \frac{x}{r}\,e^{-\lambda_0}(-\lambda_0')\eta^2rw
\{E,rw\}\,dv\,dx\\
&=&
\iint\phi'(E)e^{\mu_0}\frac{\gamma v}{\p}\cdot
\frac{x}{r}e^{-\lambda_0}\lambda_0'\eta^2rw\{E,rw\}\,dv\,dx.
\end{eqnarray*}
Using  the first identity in Lemma~\ref{identities},
\begin{eqnarray*}
V
&=&
-\gamma\iint|\phi'(E)|e^{2\mu_0-\lambda_0}\lambda_0'\frac{\eta^2rw^2}{\p}
\left(r\mu_0'\p-\p+\frac{1}{\p}\right) dv\,dx\\
&=&
-\gamma\iint|\phi'(E)|e^{2\mu_0-\lambda_0}h^2
\left(\lambda_0'\mu_0'-\frac{\lambda_0'}{r}+
\frac{\lambda_0'}{r\p^2}\right) dv\,dx.
\end{eqnarray*}
By the second identity in Lemma~\ref{identities}, 
\[
W=\gamma\iint |\phi'(E)|e^{2\mu_0-\lambda_0}h^2
\left(\mu_0''+\frac{\mu_0'}{r}+\frac{2\mu_0'}{r\p^2}\right) dv\,dx,
\]
and hence
\begin{equation}\label{v+w}
V+W=\gamma\iint|\phi'(E)|e^{2\mu_0-\lambda_0}h^2
\left(\mu_0''-\lambda_0'\mu_0'+\frac{\mu_0'+\lambda_0'}{r}
+\frac{2\mu_0'-\lambda_0'}{r\p^2}\right) dv\,dx .
\end{equation}
In order to bound the expression $\A_{12}$
from below, we use the formula (\ref{ldynaclambda}) for
$\delta\lambda$ and Lemma~\ref{auxlem}:
\begin{eqnarray}
&&
-\frac{1}{\gamma}\int_0^{\infty}e^{\mu_0-\lambda_0}
(2r\mu_0'+1)(\delta\lambda)^2 dr
\nonumber \\
&&
\qquad
=-\frac{1}{\gamma}\int_0^{\infty}e^{\mu_0-\lambda_0}(2r\mu_0'+1)
16\pi^2r^2\gamma^4e^{2\mu_0+2\lambda_0}
\left(\int \phi'(E)hw\,dv\right)^2 dr\nonumber \\
&&
\qquad
\geq-4\pi\gamma^3\int e^{3\mu_0+\lambda_0}(2r\mu_0'+1)
\frac{e^{-2\lambda_0-\mu_0}}{4\pi\gamma^2 r}
\left(\lambda_0'+\mu_0'\right)\int|\phi'(E)|h^2 dv\,dx\nonumber \\
&&
\qquad
=-\gamma\iint|\phi'(E)|e^{2\mu_0-\lambda_0}h^2
\left(2\mu_0'\lambda_0'+2(\mu_0')^2+\frac{\mu_0'+\lambda_0'}{r}\right) dv\,dx .
\qquad \quad \label{lowerbound}
\end{eqnarray}
>From (\ref{v+w}) and (\ref{lowerbound}) we have
\begin{eqnarray}
&&
V+W
-\frac{1}{\gamma}
\int_0^{\infty}e^{\mu_0-\lambda_0}(2r\mu_0'+1)(\delta\lambda)^2 dr
\nonumber \\
&&
\qquad \geq
\gamma\iint|\phi'(E)|e^{2\mu_0-\lambda_0}h^2
\left[\mu_0''-3\mu_0'\lambda_0'-2(\mu_0')^2
+\frac{2\mu_0'-\lambda_0'}{r\p^2}\right].\qquad \label{lowerbound1}
\end{eqnarray}
In order to estimate the term in the rectangular brackets on the right-hand 
side of~(\ref{lowerbound1}), we express $\mu_0''$ via the field 
equation~(\ref{ee2ndo}) for the steady state and obtain,
using the fact that the source term $q_0$ in the latter equation
is non-negative,
\begin{eqnarray*}
[\dots]
&=&
4\pi\gamma^2q_0e^{2\lambda_0}-2\mu_0'\lambda_0'-3(\mu_0')^2
-\frac{\mu_0'}{r}+\frac{\lambda_0'}{r}+2\frac{\mu_0'}{r\p^2}-
\frac{\lambda_0'}{r\p^2}\\
&\geq&
-2\mu_0'(\mu_0'+\lambda_0')-(\mu_0')^2
+\frac{\mu_0'}{r}\left(\frac{2}{\p^2}-1\right)
+\frac{\lambda_0'}{r} \left(1-\frac{1}{\p^2}\right)\\
&=&
-8\pi\gamma^2re^{2\lambda_0}\nu_0'(\rho_0+\gamma p_0)
-\gamma^2(\nu_0')^2+\frac{\gamma\nu_0'}{r}\frac{1-\gamma |v|^2}{\p^2}
+ \frac{\lambda_0'}{r}\frac{\gamma |v|^2}{\p^2}.
\end{eqnarray*}
The observation that
\[
\lambda_0'=\lambda_0'+\mu_0'-\mu_0'
=\gamma\left(4\pi r(\rho_0+\gamma p_0)e^{2\lambda_0}-\nu_0'\right)
\geq -\gamma \nu_0'
\]
implies that
\begin{eqnarray*}
[\dots]
&\geq&\gamma\frac{\nu_0'}{r\p^2}
-\gamma^2\left(8\pi r\nu_0'e^{2\lambda_0}(\rho_0+\gamma p_0) 
+ (\nu_0')^2+2 \frac{\nu_0'|v|^2}{r\p^2}\right)\\
&=&
\gamma\frac{\nu_0'}{r\p^2}
\left[
1 - \gamma \left(8\pi r^2 \p^2 e^{2\lambda_0} (\rho_0+\gamma p_0) 
+ r\p^2 \nu_0' + 2 |v|^2\right)\right].
\end{eqnarray*}
Using Proposition~\ref{ssprop} there exists a constant 
$\gamma_1\in]0,\gamma_0]$ such that 
for all $0<\gamma \leq\gamma_1$,
\begin{equation}\label{smallness}
[\dots]\geq\frac{\gamma}{4}\frac{\nu_0'}{r}.
\end{equation}
Therefore, from~(\ref{u+v+w}) and (\ref{lowerbound}), 
assuming this smallness condition for $\gamma$ we obtain
the crucial estimate
\begin{eqnarray}
\A_1(\delta f)
&\geq &
\iint |\phi'(E)| \left(e^{-\lambda_0}(rw)^2|\{E,\eta\}|^2
+\frac{\gamma^2}{4} e^{2\mu_0-\lambda_0}\frac{\nu_0'}{r}h^2\right) dv\,dx
\nonumber \\
&\geq& 
C \iint |\phi'(E)|\left((rw)^2|\{E,\eta\}|^2
+\gamma^2 h^2\right) dv\,dx,
\label{crucial}
\end{eqnarray}
where we again used the bounds from Proposition~\ref{ssprop}.

\smallskip\noindent
{\em Step 3: The term $\A_2$.} 
Since this term is of higher order in $\gamma$, we expect to be able to 
prove that it is small compared to $\A_1$ for suitably small $\gamma$.
Using the decomposition~(\ref{decomp1}) and keeping in mind 
that $\eta=h/rw$ and the formula (\ref{ldynaclambda}) for $\delta\lambda$
we can rewrite the part $\A_{21}$ as follows:
\begin{eqnarray*}
\A_{21}
&=&
-8\pi\gamma^3\iint|\phi'(E)|e^{2\mu_0+\lambda_0}
\frac{rw^2}{\p}\{E,h\}\left(\int \phi'(E)h\tilde w\,d\tilde v\right) dv\,dx\\
&=&
-8\pi\gamma^3\iint|\phi'(E)|e^{2\mu_0+\lambda_0}\frac{r^2w^3}{\p}
\{E,\eta\}\left(\int \phi'(E)h\tilde w\,d\tilde v\right) dv\,dx\\
&&
{}-8\pi\gamma^3\iint|\phi'(E)|e^{2\mu_0+\lambda_0}
\frac{w}{\p}\{E,rw\} h \left(\int \phi'(E)h\tilde w\,d\tilde v\right) dv\,dx\\
&=:&
X+Y.
\end{eqnarray*}
By Lemma~\ref{auxlem},
\begin{equation}\label{helpful}
\left(\int|\phi'(E)|wh\,dv\right)^2
\leq\frac{e^{-\mu_0}}{\gamma}
\left(\rho_0+\gamma p_0\right)\int|\phi'(E)|h^2\,dv
\leq\frac{C}{\gamma}\int|\phi'(E)|h^2\,dv.
\end{equation}
In addition, Lemma~\ref{intphipr} together with the 
bounds from Proposition~\ref{ssprop} imply that
\begin{equation}\label{helpful2}
\sup_{x\in \R^3}\int |\phi'(E)| w^2 dv \leq \frac{C}{\gamma}.
\end{equation}
The bounds from Proposition~\ref{ssprop},
the Cauchy-Schwarz inequality, and the estimates
(\ref{helpful}), (\ref{helpful2}) imply that 
\begin{eqnarray*}
|X|
&\leq& 
C \gamma^3\int\left|\int |\phi'(E)|^{1/2}w 
|\phi'(E)|^{1/2}r w \{E,\eta\}\,dv\right|
\left|\int |\phi'(E)| h w \,dv\right|\,dx \\
&\leq&
C \gamma^{5/2}
\left(\iint |\phi'(E)| |rw\{E,\eta\}|^2 dv\,dx\right)^{1/2}
\left(\frac{1}{\gamma}\iint|\phi'(E)| h^2 dv\,dx\right)^{1/2}.
\end{eqnarray*}
Hence by (\ref{crucial}),
\[
|X| \leq C \gamma \A_1.
\]
In order to estimate the term $Y$ we rewrite
the first identity of Lemma~\ref{identities}:
\[
\{E,rw\}=e^{\mu_0}r\gamma\nu_0'\p+e^{\mu_0}\left(-\p+\frac{1}{\p}\right)
=\gamma e^{\mu_0}\left(r\nu_0'-\frac{v^2}{\p}\right).
\]
Thus, 
\[
\sup_{\supp f_0}|\{E,rw\}| \leq C\gamma.
\]
Using this together with the estimates from Proposition~\ref{ssprop}
we proceed as above to find that
\begin{eqnarray*}
|Y|
&\leq&
C\gamma^{7/2}
\left(\iint |\phi'(E)| h^2\,dv\,dx\right)^{1/2}
\left(\frac{1}{\gamma}\iint|\phi'(E)| h^2 \,dv\,dx\right)^{1/2} \\
&\leq&
C \gamma \A_1.
\end{eqnarray*}
>From the above estimates for $|X|$ and $|Y|$
it follows that
\[
\A_2\geq -|X|-|Y| \geq -C \gamma \A_1.
\]
>From this estimate and~(\ref{a1a2}), we finally infer that
\[
\A \geq \frac{1}{2}\A_1-C \gamma \A_1 \geq \frac{1}{4}\A_1
\]
provided $\gamma$ is sufficiently 
small. In view of (\ref{crucial}) the proof is complete. \prfe

\noindent
{\bf Remark.} The estimate (\ref{crucial}) shows that
an equivalent way of stating our
coercivity estimate is
\[
\A(\delta f)
\geq 
\frac{1}{2}
\iint |\phi'(E)| \left(e^{-\lambda_0}(rw)^2|\{E,\eta\}|^2
+\frac{\gamma^2}{4} e^{2\mu_0-\lambda_0}\frac{\nu_0'}{r}h^2\right) dv\,dx, 
\]
which should be compared with the coercivity estimate
in the Newtonian case, cf.\ \cite[Lemma~1.1]{GuRe}.

\smallskip

If we want to use Theorem~\ref{th:coercivity} to deduce a stability
result the restriction that $h$ is odd in $v$ is a problem
because the generating functions of our perturbations need
not be odd and more importantly, even if they were this property
is not preserved under the linearized flow. However, the restriction
is easily removed. To this end, we define for a function
$h=h(x,v)$ its even and odd parts with respect to $v$
as usual by
\[
h_+(x,v):=\frac{1}{2}\left(h(x,v)+h(x,-v)\right),\
h_-(x,v):=\frac{1}{2}\left(h(x,v)-h(x,-v)\right).
\]
\begin{cor}\label{cor:coercivity}
For any
spherically symmetric function $h\in C^2(\R^6)$ the estimate
\begin{eqnarray*}
\mathcal{A}(\delta f)
&\geq& 
C^\ast\iint|\phi'(E)|\,\left((rw)^2 
\left|\left\{E,\frac{h_-}{rw}\right\}\right|^2 
+ \gamma^2 |h_-|^2\right) dv\,dx \\
&&
{}+
\frac{1}{2}
\iint e^{-\lambda_0}|\phi'(E)|\,\left|\{E,h_+\}\right|^2dv\,dx
\end{eqnarray*}
holds. Here $\delta f$ is the dynamically accessible perturbation 
generated by $h$,
$C^\ast >0$ and $\gamma^\ast>0$ are as in 
Theorem~\ref{th:coercivity}, and $0<\gamma\leq\gamma^\ast$.
\end{cor}
{\bf Proof.}
We split $h$ into its even and odd parts, $h=h_+ + h_-$. Then 
\[
\delta\lambda=4\pi r \gamma^2 e^{\mu_0+\lambda_0}
\int \phi'(E)\,h_-(x,v)\,w\,dv,
\]
since $h_+$ does not contribute to the last integral.
Since $f_0$ is even in $v$ this implies that
\begin{eqnarray*}
\delta f_-
&=&
e^{-\lambda_0}\{h_+,f_0\},\\
\delta f_+
&=&
e^{-\lambda_0}\{h_-,f_0\} + 
\gamma e^{\mu_0} \phi'(E) \frac{w^2}{\p} \delta\lambda,
\end{eqnarray*}
in particular, the even part of $\delta f$ is the 
dynamically accessible perturbation induced by the odd
part of $h$. 
Hence
\begin{eqnarray*}
\A (\delta f)
&=&
\A (\delta f_+) + 
\iint e^{\lambda_0}\frac{\delta f_+ \, \delta f_-}{|\phi'(E)|}\,dv\,dx
+ \frac{1}{2}
\iint e^{\lambda_0}\frac{|\delta f_-|^2}{|\phi'(E)|}\,dv\,dx\\
&=&
\A (\delta f_+) +
\frac{1}{2}
\iint e^{-\lambda_0} |\phi'(E)|\, \left|\{E,h_+\}\right|^2dv\,dx,
\end{eqnarray*}
since the integrand of the mixed term is odd in $v$.
The assertion follows if we now apply Theorem~\ref{th:coercivity}.
\prfe 
\section{The linearized Einstein-Vlasov system}
\label{sect_lin}
\setcounter{equation}{0}
In order to linearize the spherically symmetric Einstein-Vlasov
system (\ref{vlasov})--(\ref{jdef}) about a given steady state
$(f_0,\lambda_0,\mu_0)$ we write 
\[
f(t)=f_0 +\delta f(t),\ 
\lambda(t)=\lambda_0 + \delta\lambda(t),\ 
\mu(t)=\mu_0 + \delta\mu(t),
\]
substitute this into the system, use the fact that $(f_0,\lambda_0,\mu_0)$
is a solution, and drop all terms beyond the linear ones in
$(\delta f,\delta\lambda,\delta\mu)$.
At the moment it may not be obvious that this notation is consistent
with the one from the previous sections, because the $\delta$ terms
now have a different meaning. But it turns out below that the
notation is indeed consistent, and we obtain the following system:
\begin{equation}\label{linvlasov}
\partial_t\delta f + \frac{1}{\gamma}e^{-\lambda_0}\{\delta f,E\}
-\left(\gamma \,\dot{\delta \lambda}\, w 
+ e^{\mu_0-\lambda_0}\delta\mu'\p \right)
\phi'(E)\frac{e^{\mu_0}w}{\p}=0,
\end{equation}
\begin{eqnarray}
e^{-2\lambda_0}(r\delta\lambda'-\delta\lambda\,(2r\lambda_0'-1))
&=&
4\pi \gamma r^2\delta\rho, \label{lineelambda}\\
e^{-2\lambda_0}(r\delta\mu'-\delta\lambda\,(2r\mu_0'+1))
&=&
4\pi \gamma^2 r^2\delta p,\label{lineemu}
\end{eqnarray}
where
\begin{eqnarray}
\delta\rho(t,r)
&=&
\int \p \delta f(t,x,v)\,dv, \label{linrhodef}\\
\delta p(t,r)
&=&
\int \frac{w^2}{\p }\delta f(t,x,v)\,dv.\label{linpdef}
\end{eqnarray}
Since the condition of a regular center implies that
$\delta\lambda(t,0)=0$ the quantity $\delta\lambda$ is determined by
(\ref{lineelambda}) as
\begin{equation} \label{barlambdadef}
\delta\lambda(t,r) = 
\gamma e^{2 \lambda_0}\frac{4 \pi}{r} \int_0^r s^2 \delta\rho (t,s)\,ds,
\end{equation}
which agrees with the previous definition of
$\delta\lambda$ in (\ref{varlambdadef}).
The question arises whether the linearized version of
(\ref{eelambdad}) follows from the linearized system stated
above; we will make no use of the linearized version of (\ref{ee2ndo})
to which this question of course applies as well. Indeed,
\begin{equation}
\dot{\delta\lambda} =
-4\pi \gamma r e^{\mu_0+\lambda_0}\delta\jmath,\label{lineelambdad}
\end{equation}
where
\begin{equation}
\delta\jmath(t,r) =
\int  w \,\delta f(t,x,v)\,dv. \label{linjdef}
\end{equation}
To see this we observe first that by (\ref{barlambdadef}),
\[
\dot{\delta\lambda}(t,r) = 
\gamma e^{2 \lambda_0}\frac{4 \pi}{r} 
\int_0^r s^2 \partial_t\delta\rho (t,s)\,ds.
\]
By (\ref{linvlasov}) and integration by parts,
\[
\partial_t\delta\rho
= -e^{\mu_0 -\lambda_0} \int v\cdot\partial_x \delta f\, dv 
- 2 \mu_0'e^{\mu_0 -\lambda_0}\int w \,\delta f\, dv - 
\dot{\delta\lambda}\, (\rho_0 + \gamma p_0).
\]
If we substitute this into the equation for $\dot{\delta\lambda}$ and integrate
the first term by parts with respect to $x$ we find that
\[
r \dot{\delta\lambda} + 4 \pi r^2 \gamma e^{\mu_0+\lambda_0}\delta\jmath
= - 4 \pi \gamma e^{2 \lambda_0} \int_0^r \left(s \dot{\delta\lambda} + 
4 \pi s^2 \gamma e^{\mu_0+\lambda_0}\delta\jmath\right)
(\rho_0 +\gamma p_0)\,s\, ds.
\]
Since the term in parenthesis vanishes at the origin it vanishes
everywhere, which is the assertion.

Since we are restricting our linear stability theory to the 
class of dynamically accessible perturbations,
we limit ourselves to an existence theorem for such solutions. 
In particular,
we show the essential fact 
that linearly dynamically accessible data retain this
property under the flow of the linearized system.

In order to avoid purely technical complications we
assume from now on that in addition to the previous
assumptions, $\Phi \in C^2(\R)$. We comment on this assumption 
in a remark at the end of this section.
\begin{theorem}\label{th:ex}
Let $\open{h}\in C^2(\R^6)$ be a spherically symmetric function 
which according to (\ref{ldynacdef}) generates a
linearly dynamically accessible perturbation $\open{\delta f}$.
Then there exists a unique solution 
$\delta f\in C^{1}([0,\infty[\times \R^6)$ to the 
linearized Einstein-Vlasov system (\ref{linvlasov})--(\ref{linpdef})
with $\delta f(0)=\open{\delta f}$. Furthermore, there exists 
$h\in C^{1,2}([0,\infty[\times \R^6)$ such that 
\be\label{barfdef}
\delta f(t)=e^{-\lambda_0}\{h(t),f_0\}
+4\pi\gamma^3re^{2 \mu_0+\lambda_0}\phi'(E)\frac{w^2}{\p}
\int \phi'(E)h\tilde w\,d\tilde v,
\ee
i.e., $\delta f(t)$ is linearly dynamically accessible.
The generating function $h$ is the unique solution to the 
transport equation
\be\label{eq:heqn}
\partial_t h + \frac{1}{\gamma}e^{-\lambda_0}\{h,E\} +
e^{\mu_0} \delta\lambda  \frac{w^2}{\p } +
\frac{1}{\gamma} E \delta\mu =0
\ee
with initial value $h(0)=\open{h}$.
\end{theorem}
{\bf Proof.}
The proof proceeds as follows. First we establish existence, 
uniqueness, and regularity of the solution $h$ to the 
equation (\ref{eq:heqn}),
where $\delta\lambda$ and $\delta\mu$ are defined via the field 
equations (\ref{lineelambda}) and (\ref{lineemu}),
with source terms induced by $\delta f$ as defined in~(\ref{barfdef}). 
Then we prove that $(\delta f,\delta\lambda,\delta\mu)$ 
indeed solves the linearized Einstein-Vlasov system.

Equation (\ref{eq:heqn}) is a first order inhomogeneous transport 
equation and its characteristics
$s\mapsto (X(s,t,x,v),V(s,t,x,v))$ are defined as the solutions of
\begin{eqnarray*}
\dot x
&=&
e^{\mu_0(x)-\lambda_0(x)}\frac{v}{\p},\\
\dot v
&=&
-\frac{1}{\gamma}e^{\mu_0(x)-\lambda_0(x)}\nabla \mu_0 (x)\p
\end{eqnarray*}
with initial condition
\[
X(t,t,x,v)=x,\ V(t,t,x,v)=v.
\]
The metric coefficients of the steady state
can of course be viewed as functions on $\R^3$,
and as such, $\lambda_0\in C^2(\R^3)$ and $\mu_0\in C^3(\R^3)$,
cf.\ \cite{RR93,RR00}. Using the abbreviations 
$z=(x,v)$ and $Z=(X,V)$ we see that 
$Z(s,t,\cdot):\R^6\to\R^6$ is a $C^2$ diffeomorphism.
Now assume that $h$ is a solution to (\ref{eq:heqn}).
Then integration along the characteristics implies that
\begin{equation}\label{integro}
h(t,z)=\open{h}\,(Z(0,t,z))-
\int_0^t\left(e^{\mu_0}\delta\lambda  \frac{w^2}{\p } +
\frac{1}{\gamma} E \delta\mu\right)(s,Z(s,t,z))\,ds.
\end{equation}
We wish to find a solution to this equation, where
$\delta\lambda$ and $\delta\mu$ are given as the solutions 
to the field equations
(\ref{lineelambda}) and (\ref{lineemu}) with source terms induced
by $\delta f$ which in turn is defined via (\ref{barfdef}).

To construct solutions to (\ref{integro}), we apply a simple iteration scheme.
For any spherically symmetric
function $g\in C^2(\R^6)$ we define
the function $\delta f_g$ by (\ref{barfdef}) with $h$ replaced by $g$.
Clearly, $\delta f_g\in C^1(\R^6)$, and $\delta f_g$ is supported
in the support of $f_0$. The induced source terms $\delta\rho_g$ 
and $\delta p_g$ have the same regularity and are compactly 
supported as well. The equation (\ref{barlambdadef}) shows
that the induced metric component $\delta\lambda_g \in C^2(\R^3)$.
Moreover, the formula in Proposition~\ref{ldynacprop} shows
that $\delta\lambda_g$ is also compactly supported.
Next we can define $\delta\mu'_g$ by (\ref{lineemu}) and using
the boundary condition $\delta\mu_g(\infty)= 0$ we find that 
$\delta\mu_g \in C^2(\R^3)$ is compactly supported as well.
Moreover,
\begin{equation} \label{barestchain}
||\delta\lambda_{g}||_{C^2_b}+||\delta\mu_{g}||_{C^2_b}
\leq C \left(||\delta\rho_{g}||_{C^1_b}+||\delta p_{g}||_{C^1_b}\right)
\leq C ||\delta f_{g}||_{C^1_b} \leq C ||g||_{C^2_b}
\end{equation}
where the constant depends on the given steady state
and the norms extend only over the support of
the steady state which is compact.
Suppose now that $g\in C^{1,2}([0,\infty[\times \R^6)$
so that $g(t) \in C^2(\R^6)$ for $t\geq 0$.
Motivated by (\ref{integro}) we define
\begin{equation}\label{T}
(Tg)(t,z):=\open{h}\,(Z(0,t,z))-
\int_0^t\left(e^{\mu_0}\delta\lambda_{g(s)}\frac{w^2}{\p}
+\frac{1}{\gamma}E\delta\mu_{g(s)}\right)(Z(s,t,z))\,ds.
\end{equation}
It is straight forward to see that
$Tg\in C^{1,2}([0,\infty[\times \R^6)$, and the
linearity of the problem together with (\ref{barestchain})
imply that 
for $g_1,g_2\in C^{1,2}([0,\infty[,\R^6)$  we obtain the estimate
\begin{equation}\label{bound1}
\|Tg_1(t)-Tg_2(t)\|_{C^2_b}\leq C\int_0^t\|g_1(s)-g_2(s)\|_{C^2_b}.
\end{equation}
Now let 
\[
h_0(t,z):=\open{h}\,(z),\ h_{n+1}:=Th_n.
\]
The estimate (\ref{bound1}) implies that 
$h_n$ converges $t$-locally uniformly  
to some $h \in C^{1,2}([0,\infty[\times \R^6)$. 
Using again the estimates in (\ref{barestchain}) and linearity
it follows that $\delta f_{h_n}$ converges $t$-locally 
uniformly to $\delta f_h\in C^1([0,\infty[\times \R^6)$ and
$\delta\lambda_{h_n}$, $\delta\mu_{h_n}$ converge $t$-locally uniformly to 
$\delta\lambda_h, \delta\mu_h\in C^{1,2}([0,\infty[\times \R^6)$.
In particular, $h$ solves (\ref{integro}) and hence also (\ref{eq:heqn}). 
The uniqueness of the solution is clear. 
 
In order to show that $(\delta f,\delta\lambda,\delta\mu)$ 
solves the linearized Einstein-Vlasov system, it only remains to check 
the Vlasov equation~(\ref{linvlasov})
as the two field equations (\ref{lineelambda}) and 
(\ref{lineemu}) hold by definition of $\delta\lambda$ and $\delta\mu$.
The definition of $\delta f$ in terms of $h$ implies that
\[
\partial_t \delta f = e^{-\lambda_0}\{\partial_t h,f_0\}
+ e^{\mu_0} \gamma \dot{\delta\lambda}\phi'(E) \frac{w^2}{\p }.
\]
Comparing this with (\ref{linvlasov}) we see that the latter equation 
becomes equivalent to the relation
\begin{eqnarray*}
\{\partial_t h, f_0\}
&=&
-\frac{1}{\gamma}\{\delta f,E\} - e^{2\mu_0}\delta\mu' w\, \phi'(E)\\
&=&
-\frac{1}{\gamma}\{e^{-\lambda_0}\{h,f_0\},E\} - 
\left\{ e^{\mu_0}\delta\lambda\, \phi'(E)\frac{w^2}{\p} ,E\right\}
- e^{2\mu_0}\delta\mu' w\, \phi'(E).
\end{eqnarray*}
The fact that $f_0=\phi(E)$ and $\{\phi'(E),E\}=0= \{E,E\}$
together with the product rule (\ref{productrule})
imply that this relation is again equivalent to
\begin{eqnarray*}
\{\partial_t h, f_0\}
&=&
-\frac{1}{\gamma} \{e^{-\lambda_0} \phi'(E)\{h, E\},E\}  
- \left\{e^{\mu_0}\delta\lambda\, \phi'(E)\frac{w^2}{\p} ,E\right\} \\
&&
{} - e^{2\mu_0}\delta\mu' w\, \phi'(E)\\
&=&
-\frac{1}{\gamma} \phi'(E) \{e^{-\lambda_0} \{h, E\},E\} 
- \phi'(E) \left\{e^{\mu_0}\delta\lambda  \frac{w^2}{\p} ,E\right\}\\
&&
{} - \frac{1}{\gamma} \phi'(E)\{E \delta\mu,E\},
\end{eqnarray*}
i.e.
\[
\left\{\partial_t h + \frac{1}{\gamma}e^{-\lambda_0} \{h, E\}
+ e^{\mu_0} \delta\lambda  \frac{w^2}{\p }
+\frac{1}{\gamma}E \delta\mu,f_0\right\}=0
\]
which is implied by (\ref{eq:heqn}).
\prfe
{\bf Remark.}
We emphasize that a global existence and uniqueness result for 
general smooth data holds as well. 
The proof follows a simple iteration scheme analogous
to the corresponding result for the Vlasov-Poisson case~\cite{BMR}.\\
We also note that the values of $h$ outside the support of
$f_0$ are actually irrelevant and it would be sufficient
to consider the generating function $h$ as defined only
on this support.

\smallskip

The free energy $\A$ defined in (\ref{secvar}) is conserved
along solutions of the linearized system. This fact, which is
clearly important for our stability result, is shown next.
\begin{proposition}\label{encons}
Any linearly dynamically accessible solution 
as construct\-ed in Theorem~\ref{th:ex}
preserves the energy $\mathcal{A}$.
\end{proposition}
{\bf Remark.} The above assertion is true
for any sufficiently smooth solution of the
linearized system, provided in particular that the first
integral in $\mathcal{A}(\delta f(t))$ is defined.
As the remark at the end of Section~\ref{sect_dynac} shows,
this is the case for linearly dynamically accessible solutions
as constructed in Theorem~\ref{th:ex}. In the following
proof we make use of this structure only to guarantee
the existence of the otherwise questionable integrals.

\smallskip

\noindent
{\bf Proof of Proposition~\ref{encons}.}
Clearly,
\begin{eqnarray*}
\frac{1}{2}\frac{d}{dt}\iint e^{\lambda_0}
\frac{\delta f^2}{|\phi'(E)|}\,dv\,dx
&=&
\iint e^{\lambda_0}\frac{\delta f}{|\phi'(E)|}\partial_t \delta f\,dv\,dx\\
&=&
\iint \frac{\delta f }{\gamma \phi'(E)} \{\delta f,E\}\,dv\,dx\\
&&
{}-\iint \frac{e^{\lambda_0+\mu_0}}{\p} \delta f 
\left(\gamma \dot{\delta \lambda}w + e^{\mu_0-\lambda_0}
\delta\mu' \p\right) dv\,dx\\
&=&
\iint \frac{\delta f }{\gamma \phi'(E)} \{\delta f,E\}\,dv\,dx\\
&&
{}-\int \left(\gamma e^{\lambda_0+\mu_0}\delta p\, \dot{\delta \lambda}
+ e^{2 \mu_0}\delta\jmath\,\delta\mu' \right) dx.
\end{eqnarray*}
For the first term on the right hand side we use the product
rule (\ref{productrule}) and the identity (\ref{intbypartsimpl}) 
to conclude that
\begin{eqnarray*}
\iint\frac{\delta f}{|\phi'(E)|}\{E,\delta f\}\,dv\,dx
&=&-\frac{1}{2}\iint\frac{1}{\phi'(E)}\{E,\delta f^2\}\,dv\,dx\\
&=&\frac{1}{2}\iint \delta f^2\left\{E,\frac{1}{\phi'(E)}\right\}\,dv\,dx=0.
\end{eqnarray*}
Hence the linearized field equations (\ref{lineemu}) 
and (\ref{lineelambdad}) imply that
\begin{eqnarray*}
&&
\frac{1}{2}\frac{d}{dt}\iint e^{\lambda_0}
\frac{{\delta f}^2}{|\phi'(E)|}\,dv\,dx
=
- \int \left(\gamma e^{\lambda_0+\mu_0}\delta p\, \dot{\delta \lambda}
+ e^{2 \mu_0}\delta\jmath\,\delta\mu' \right) dx\\
&&
=
- \int_0^\infty\left(\dot{\delta \lambda}
\frac{e^{\mu_0-\lambda_0}}{\gamma r^2}
\left(r \delta\mu' -\delta\lambda\left(2 r \mu_0'+1\right)\right)
- \frac{e^{\mu_0-\lambda_0}}{\gamma r}\dot{\delta \lambda}\right)
r^2dr\\
&&
=\frac{1}{\gamma}\int_0^\infty e^{\mu_0-\lambda_0} \left(2 r \mu_0'+1\right)
\dot{\delta \lambda}\,\delta \lambda\,dr\\
&&
= \frac{1}{2 \gamma}\frac{d}{dt}
\int_0^\infty e^{\mu_0-\lambda_0} \left(2 r \mu_0'+1\right)
(\delta \lambda)^2dr
\end{eqnarray*}
as required, and the proof is complete.
\prfe
{\bf Remark.}
In Theorem~\ref{th:ex}, the assumption $\Phi\in C^2(\R)$ can be relaxed to a 
$C^1$ assumption.
The existence theory can then be developed for
$h\in C^{1}([0,\infty[\times\R^6)$. 
Since $\delta f$ is then only continuous, it becomes 
more technical to justify Proposition~\ref{encons},
but this can be done analogously to the proof of \cite[Theorem~5.1]{BMR}.
\section{Linear stability} \label{sect_linstab}
\setcounter{equation}{0}
Our coercivity estimate in the form of Corollary~\ref{cor:coercivity}
and the conservation of the free energy according to 
Proposition~\ref{encons} immediately imply the following result.
\begin{theorem}\label{th:maingamma}
Let $\Phi$ satisfy the assumptions stated above,
let $\gamma^\ast$ and $C^\ast$ be as in Theorem~\ref{th:coercivity},
and let $0<\gamma \leq \gamma^\ast$. Then the corresponding steady state
introduced in Proposition~\ref{ssprop} is linearly stable in 
the following sense: For any spherically symmetric function
$\open{h}\in C^2(\R^6)$ the solution of the linearized
Einstein-Vlasov system (\ref{linvlasov})--(\ref{linpdef})
with the dynamically accessible state $\open{\delta f}$ generated by $\open{h}$
according to (\ref{ldynacdef}) as initial datum satisfies
for all times $t\geq 0$ the estimate
\begin{eqnarray*}
C^\ast\iint|\phi'(E)|\,\left((rw)^2 
\left|\left\{E,\frac{h_-(t)}{rw}\right\}\right|^2 
+ \gamma^2 |h_-(t)|^2\right) dv\,dx 
&&\\
+
\frac{1}{2}
\iint e^{-\lambda_0} |\phi'(E)|\, \left|\{E,h_+(t)\}\right|^2dv\,dx
&\leq&
\mathcal{A}(\open{\delta f}).
\end{eqnarray*}
\end{theorem}

\noindent
{\bf Remark.} The left hand side is an acceptable
measure of the size of the linearized perturbation from the given 
steady state,
in particular, if it vanishes then so does the perturbation $\delta f(t)$.
The fact that the left hand side controls $h(t)$ only on the 
support of the steady
state is natural in a linearized approach. The right hand side
can clearly be made as small as desired by making the initial
perturbation small in a suitable sense.

\smallskip

The result above is acceptable as a linear stability result,
except for the fact that $\gamma=1/c^2$ has to be chosen small
when in a given set of units this quantity has a definite value. 
We therefore recast our result into one on the stability
of a one-parameter family of steady states of the Einstein-Vlasov system
with $\gamma=1$ by using the scaling properties
of the system. 
This will also provide a quantitative version of the Ze'ldovitch observation
which was explained in the introduction.

In order to do so we now denote the Einstein-Vlasov system
(\ref{vlasov})--(\ref{qdef}), i.e., including the parameter
$\gamma$, by (EV$_\gamma$) and by (EV) the system with $\gamma=1$.  
A straight forward computation shows that
if $f$ is a solution of (EV$_\gamma$)
with the corresponding metric coefficients $\lambda$ and $\mu$,
then
\[
T^{\gamma}f(t,x,v):=\gamma^{-3/2} f(t,\gamma^{-1/2}x, \gamma^{-1/2} v)
\]
defines a solution of (EV) with metric coefficients
\[
T^{\gamma}\lambda(t,r):=\lambda(t,\gamma^{-1/2} r),\ 
T^{\gamma}\mu(t,r):=\mu(t,\gamma^{-1/2} r).
\]
For a fixed $\phi$ as specified above, 
the steady state $(f_0,\lambda_0,\gamma_0)$ of (EV$_\gamma$)
with $0<\gamma\leq \gamma_0$---this is actually a family of steady states,
one for each (EV$_\gamma$)---is now mapped into
the one-parameter family of steady states of (EV) given by
\[
]0,\gamma_0]\ni\gamma\mapsto (f_0^{\gamma},\lambda_0^{\gamma},\mu_0^{\gamma})
=(T^{\gamma}f_0,T^{\gamma}\lambda_0,T^{\gamma}\mu_0);
\]
here $\gamma_0$ is from Proposition~\ref{ssprop}.
It should be carefully noted that all the members of this family
are steady states of (EV), i.e., of the system with $\gamma=1$,
and $\gamma\in]0,\gamma_0]$ is now the parameter which parametrizes
this family. In particular
\[
f_0^{\gamma}=\gamma^{-3/2}\phi(E)= \phi_\gamma(E)
\]
where
\[
E(x,v):=e^{\mu_0^{\gamma}(r)}\sqrt{1+|v|^2}\ 
\mbox{and}\ \phi_\gamma:=\gamma^{-3/2}\phi.
\]
In order to understand what it means to move to smaller
values of $\gamma$ in this family we observe that by construction
$\mu_0^\gamma(R^\gamma)=0$ holds for a unique radius $R^\gamma>0$
which determines the boundary of the spatial support of the steady state,
i.e., the boundary or surface of the galaxy or globular cluster.
 At the center $r=0$ we have 
$\mu_0^{\gamma}(0) = \gamma \open{\nu}$.
The expression
\[
z:= \frac{e^{\mu^\gamma_0(R^\gamma)}}{e^{\mu^{\gamma}_0(0)}} -1 =
 \frac{1}{e^{\gamma\open{\nu}}} -1 >0
\]
is the redshift of a photon which is emitted at the center
and received at the surface of the mass distribution, and it is
a measure for how strong relativistic effects in the configuration
are. Expressed in units where the speed of light
$c =1$ the above limit
$\gamma \to 0$ means that we consider steady states for which 
$z$ is close to $0$, i.e., relativistic effects are weak. 

In order to recast our stability result for the case of (EV)
we have to check how the various quantities which are involved
behave under the scaling operator $T^\gamma$.
To this end we rename the particle energy in the context
of (EV$_\gamma$) as
\[
E_\gamma = E_\gamma(x,v)=e^{\mu_0(r)}\sqrt{1+\gamma |v|^2}
= E(\gamma^{1/2}x,\gamma^{1/2}v).
\]
In order to understand the 
behavior of the Poisson bracket (\ref{pbdef}) under the scaling
operator $T^\gamma$, we define for 
a given function $h=h(x,v)$, 
\[
h^\gamma(x,v):=\gamma^{-1}h(\gamma^{1/2}x,\gamma^{1/2}v).
\]
Then the relation 
\[
\{h,f_0^{\gamma}\}(x,v)=T^{\gamma}\{h^\gamma,f_0\}
\]
holds. Next,
a simple change of variables argument implies that
\beas
&&
4\pi re^{\mu^{\gamma}_0+\lambda^{\gamma}_0}\phi_\gamma'(E)
\frac{w^2}{\p}\int \phi'_\gamma (E)\,h \tilde w\,d\tilde v\\
&&
\qquad\qquad = T^{\gamma}\left(4\pi\gamma^3 re^{\mu_0+\lambda_0}
\phi'(E_{\gamma})\frac{w^2}{\p}\int \phi'(E_{\gamma})
\,h^\gamma\tilde w\,d\tilde v\right).
\eeas
In particular, if 
\begin{equation}\label{ldynacdef1}
\delta f_h=e^{-\lambda_0^\gamma}\{h,f^{\gamma}_0\}+
4\pi re^{\mu^{\gamma}_0+\lambda^{\gamma}_0}\phi'_\gamma(E)\frac{w^2}{\p}
\int \phi'_\gamma(E)\,h\,\tilde w\, d\tilde v
\end{equation}
is the linearly dynamically accessible state for (EV)
generated by $h$, then
\be\label{eq:relation2}
\delta f_h=T^{\gamma}\delta f_{h^\gamma},
\ee
where $\delta f_{h^\gamma}$ is the associated linearly
dynamically accessible state for (EV$_\gamma$), generated
by $h^\gamma$ as 
defined in (\ref{ldynacdef}).

Finally, to understand the behavior 
of the the free energy we first must again be careful with the
notation. The free energy associated with the steady state
$(f_0,\lambda_0,\mu_0)$ of (EV$_\gamma$)
and defined in (\ref{secvar}) is still denoted by
$\A$, while the corresponding quantity associated
with the steady state
$(f_0^\gamma,\lambda_0^\gamma,\mu_0^\gamma)$ of (EV) is defined by
\[
\mathcal{A^\gamma}(\delta f):=
\frac{1}{2}\iint \frac{e^{\lambda^\gamma_0}}{|\phi_\gamma'(E)|}
(\delta f)^2 dv\,dx
- \frac{1}{2}\int_0^\infty e^{\mu^\gamma_0 - \lambda^\gamma_0}
\left(2 r \left(\mu^\gamma_0\right)' +1\right) (\delta\lambda)^2\,dr.
\]
A simple calculation shows that
\be\label{eq:relation4}
\A (\delta f)=\gamma^{-3/2}\A^\gamma (T^{\gamma}\delta f).
\ee
We now recast Theorem~\ref{th:maingamma}, which is a result for (EV$_\gamma$),
into a result for (EV).
\begin{theorem}\label{th:main}
Let $\Phi$ satisfy the assumptions stated above
and let $\gamma^\ast$ and $C^\ast$ be as in Theorem~\ref{th:coercivity}.
Then provided
$0<\gamma \leq \gamma^\ast$ the corresponding steady state
$(f_0^\gamma,\lambda_0^\gamma,\mu_0^\gamma)$ of (EV)
is linearly stable in the following sense: 
For any spherically symmetric function
$\open{h}\in C^2(\R^6)$ the solution $f_h$ of the linearized
Einstein-Vlasov system 
with the dynamically accessible state 
$\open{\delta f}$ generated by $\open{h}$
according to (\ref{ldynacdef1}) as initial datum satisfies
for all times $t\geq 0$ the estimate
\begin{eqnarray*}
C^\ast\iint|\phi_\gamma'(E)|\,\left((rw)^2 
\left|\left\{E,\frac{h_-(t)}{rw}\right\}\right|^2 
+ |h_-(t)|^2\right) dv\,dx 
&&\\
+
\frac{1}{2}
\iint e^{-\lambda_0^\gamma} |\phi_\gamma'(E)|\, \left|\{E,h_+(t)\}\right|^2dv\,dx
&\leq&
\A^\gamma(\open{\delta f}).
\end{eqnarray*}
\end{theorem}
{\bf Proof.}
In the statement of the theorem the linearized Einstein-Vlasov
system is now the system (\ref{linvlasov})--(\ref{linpdef})
with $\gamma=1$, and $\A^\gamma$ is a conserved quantity,
in particular along solutions which are linearly dynamically
accessible from the steady state $(f_0^\gamma,\lambda_0^\gamma,\mu_0^\gamma)$
about which the system (EV) was linearized. 
Hence by (\ref{eq:relation2}) and (\ref{eq:relation4}),
\[
\A^\gamma(\open{\delta f}) =
\A^\gamma(\delta f_h(t)) = \A^\gamma(T^\gamma \delta f_{h^\gamma}(t))=
\gamma^{3/2} \A(\delta f_{h^\gamma}(t)).
\]
If we apply Corollary~\ref{cor:coercivity} it follows that
\begin{eqnarray*}
\A^\gamma(\open{\delta f})
&\geq&
C^\ast\gamma^{3/2} \iint|\phi'(E_{\gamma})|\,\left((rw)^2 
\left|\left\{E_{\gamma},\frac{h_-^\gamma(t)}{rw}\right\}\right|^2 
+ \gamma^2 |h_-^\gamma(t)|^2\right) dv\,dx \\
&&
{}+
\gamma^{3/2} \frac{1}{2}
\iint e^{-\lambda_0} |\phi'(E_{\gamma})|\, \left|\{E_{\gamma},h_+^\gamma(t)\}\right|^2dv\,dx.
\end{eqnarray*}
Now we apply a change of variables to turn $h^\gamma$ into $h$
under these integrals and the result follows.
\prfe

\bigskip

\noindent
{\bf Acknowledgments.} This research was partly supported by the National Science Foundation under grant CMG-0530862
and the European Research Council under grant ERC-240385.
The second author wants to thank Markus Kunze
for his remark that a dynamical system should not change its stability
properties immediately if one turns on a small parameter.
He also thanks Phil Morrison for many discussions about
the Hamiltonian perspective, for kinetic equations in general and the
Einstein-Vlasov system in particular.


\begin{thebibliography}{AAAA}

\bibitem{And05}
{\sc Andr\'{e}asson, H.},
The Einstein-Vlasov System/Kinetic Theory.
{\em Living Rev. Relativity}\ {\bf 14}, 4.~URL (cited on 16.1.2012) (2011).

\bibitem{AKR_bh}
{\sc Andr\'{e}asson, H., Kunze, M., Rein, G.},
The formation of black holes in spherically symmetric gravitational collapse.
{\em Mathematische Annalen} {\bf 350}, 683--705 (2011).

\bibitem{AKR}
{\sc Andr\'{e}asson, H., Kunze, M., Rein, G.},
Existence of axially symmetric static solutions of the 
Einstein-Vlasov system.
{\em Commun. Math. Phys.}
\ {\bf 308}, 23--47 (2011).

\bibitem{AnRe1}
{\sc Andr\'{e}asson, H., Rein, G.},
A numerical investigation of the stability of steady states 
and critical phenomena for the spherically symmetric 
Einstein-Vlasov system.
{\em Class.\ Quantum Grav.}\ 
{\bf 23}, 3659--3677 (2006).

\bibitem{BMR}
{\sc Batt, J., Morrison, P., Rein, G.},
Linear stability of stationary solutions of the Vlasov-Poisson system
in three dimensions.
{\em Arch.\ Rational Mech.\ Anal.}\
{\bf 130}, 163--182 (1995).

\bibitem{DR}
{\sc Dafermos, M., Rendall, A.},
An extension principle for Einstein-Vlasov system in spherical symmetry.
{\em Ann.\ de l'Inst.\ H.\ Poincar\'e}
{\bf 6} 1137--1155 (2005).

\bibitem{Gu}
{\sc Guo, Y.},
Variational method in polytropic galaxies.
{\em Arch.\ Rational Mech.\ Anal.}\
{\bf 150}, 209--224 (1999).

\bibitem{GuRe01}
{\sc Guo, Y., Rein, G.},
Isotropic steady states in galactic dynamics.
{\em Commun. Math. Phys.}\
{\bf 219}, 607--629 (2001).

\bibitem{GuRe}
{\sc Guo, Y., Rein, G.},
A non-variational approach to nonlinear stability in stellar dynamics 
applied to the King model.
{\em Commun. Math. Phys.}\
{\bf 271}, 489--509 (2007). 

\bibitem{HaRe}
{\sc Had\v zi\'c, M., Rein, G.},
Global existence and nonlinear stability for the relativistic 
Vlasov-Poisson system in the gravitational case. 
{\em Indiana Univ. Math. J.}\
{\bf 56}, 2453--2488 (2007).

\bibitem{IT68}
{\sc Ipser, J., Thorne, K.},
Relativistic, spherically symmetric star clusters
I.\ Stability theory for radial perturbations.
{\em Astrophys.\ J.}\
{\bf 154}, 251--270 (1968).

\bibitem{KM}
{\sc Kandrup, H., Morrison, P.},
Hamiltonian structure of the Vlasov-Einstein system and the problem 
of stability for spherical relativistic star clusters.
{\em Annals of Physics}\
{\bf 225}, 114--166 (1993).

\bibitem{KS} 
{\sc Kandrup, H., Sygnet, J. ~F.}, 
A simple proof of dynamical stability for a class of spherical clusters. 
{\em Astrophys.~J.}\
{\bf 298}, 27--33 (1985).

\bibitem{LeMeRa2}
{\sc Lemou, M., Mehats, F., Rapha\"el, P.},
Stable ground states for the relativistic gravitational 
Vlasov-Poisson system.
{\em Commun.\ Partial Differential Eqns.}\
{\bf 34}, 703--721 (2009).

\bibitem{LeMeRa1}
{\sc Lemou, M., Mehats, F., Rapha\"el, P.},
A new variational approach to the stability of gravitational systems.
{\em Commun.\  Math.\  Phys.}
{\bf 302}, 161--224 (2011).

\bibitem{LeMeRa3}
{\sc Lemou, M., Mehats, F., Rapha\"el, P.},
Orbital stability of spherical systems. 
{\em Inventiones Math.}\
{\bf 187}, 145--194 (2012).

\bibitem{Mou}
{\sc Mouhot, C.},
Stabilit\'{e} orbitale pour le syst\`{e}me de Vlasov-Poisson gravitationnel, 
d'apr\`{e}s Lemou-M\'{e}hats-Rapha\"el, Guo, Lin, Rein et al.
{\em S\'{e}minaire Nicolas Bourbaki Nov.~2011}, arXiv:1201.2275 (2012).

\bibitem{Rein94}
{\sc Rein, G.},
Static solutions of the spherically symmetric Vlasov-Einstein
system.
{\em Math.\ Proc.\ Camb.\ Phil.\ Soc.}\ 
{\bf 115}, 559--570 (1994).

\bibitem{Rein95}
{\sc Rein, G.},
{\em The Vlasov-Einstein System with Surface Symmetry},
Ha\-bi\-li\-tations\-schrift, M\"unchen 1995.

\bibitem{Rein07}
{\sc Rein, G.},
Collisionless kinetic equations from astrophysics---The Vlasov-Poisson
system.
In {\em Handbook of Differential Equations, Evolutionary Equations,
vol.~3}, edited by C.~M.~Dafermos and E.~Feireisl,
Elsevier (2007).

\bibitem{RR92a}
{\sc Rein, G., Rendall, A.},
Global existence of solutions of the spherically symmetric 
Vlasov-Einstein system with small initial data,
{\em Commun.\ Math.\ Phys.}\ {\bf 150}, 561--583 (1992).
Erratum: {\em Commun.\ Math.\ Phys.}\ {\bf 176}, 475--478 (1996).

\bibitem{RR92b}
{\sc G.~Rein and A.~Rendall}.
The Newtomian limit of the spherically symmetric Vlasov-Einstein system.
{\em Commun.\ Math.\ Phys.}\
{\bf 150}, 585--591 (1992).  

\bibitem{RR93}
{\sc Rein, G., Rendall, A.},
Smooth static solutions of the
spherically symmetric Vlasov-Einstein system.
{\em Ann.\ de l'Inst.\ H.\ Poincar\'e, Physique Th\'eorique}
{\bf 59}, 383--397 (1993).

\bibitem{RR00}
{\sc Rein, G., Rendall, A.},
Compact support of spherically symmetric equilibria in
non-relativistic and relativistic galactic dynamics.
{\em Math.\ Proc.\ Camb.\ Phil.\ Soc.}\ 
{\bf 128}, 363--380 (2000).

\bibitem{Wo}
{\sc Wolansky, G.},
Static solutions of the Vlasov-Einstein system.
{\em Arch.\ Rational Mech.\ Anal.}\ {\bf 156} 205--230 (2001).

\bibitem{ZeNo}
{\sc Zel'dovich, Ya. B. , Novikov, I. D.},
{\em Relativistic Astrophysics} 
{\bf Vol. 1}, 
Chicago: Chicago University Press (1971).

\bibitem{ZePo}
{\sc Zel'dovich, Ya. B. , Podurets, M. A.}, 
The evolution of a system of gravitationally interacting point masses.
{\em Soviet Astronomy -- AJ} 
{\bf 9}, 742Ð749 (1965), 
translated from {\em Astronomicheskii Zhurnal}~{\bf 42}. 

\end{thebibliography}
\end{document}